\documentclass{aa}

\pdfoutput=1

\usepackage{graphicx}
\usepackage{txfonts}
\usepackage{multirow}
\usepackage{subfigure}
\usepackage{xspace}
\usepackage[switch]{lineno}
\usepackage{hyperref}

\newcommand{\hess}{H.E.S.S.\xspace}
\newcommand{\ct}{\emph{ctools}\xspace}
\newcommand{\gp}{\emph{Gammapy}\xspace}
\newcommand{\hap}{\emph{HAP}\xspace}
\newcommand{\rxj}{RX~J1713.7-3946\xspace}
\newcommand{\msh}{MSH~15-5\emph{2}\xspace}
\newcommand{\pks}{PKS~2155-304\xspace}
\newcommand{\crab}{Crab nebula\xspace}
\newcommand{\gam}{$\gamma$\xspace}
\newcommand{\psimax}{$\Psi_\mathrm{max}$\xspace}
\newcommand{\ethr}{$E_\mathrm{thr}$\xspace}

\makeatletter
\renewcommand*\aa@pageof{, page \thepage{} of \pageref*{LastPage}}
\makeatother

\begin{document}

\title{Validation of open-source science tools and background model construction in \gam-ray astronomy}

\author{L. Mohrmann \and
        A. Specovius \and
        D. Tiziani \and
        S. Funk \and
        D. Malyshev \and
        K. Nakashima \and
        C. van Eldik
}
\authorrunning{L. Mohrmann et al.}

\institute{Erlangen Centre for Astroparticle Physics, University of Erlangen-Nuremberg,
           Erwin-Rommel-Str. 1, 91058 Erlangen, Germany\\
           \email{lars.mohrmann@fau.de}
}

\date{Accepted for publication in A\&A}

\abstract
{
In classical analyses of \gam-ray data from imaging atmospheric Cherenkov telescopes (IACTs), such as the High Energy Stereoscopic System (\hess), aperture photometry, or photon counting, is applied in a (typically circular) region of interest (RoI) encompassing the source.
A key element in the analysis is to estimate the amount of background in the RoI due to residual cosmic ray-induced air showers in the data.
Various standard background estimation techniques have been developed in the last decades, most of them rely on a measurement of the background from source-free regions within the observed field of view.
However, in particular in the Galactic plane, source analysis and background estimation are hampered by the large number of, sometimes overlapping, \gam-ray sources and large-scale diffuse \gam-ray emission.

For complicated fields of view, a three-dimensional (3D) likelihood analysis shows the potential to be superior to classical analysis.
In this analysis technique, a spectromorphological model, consisting of one or multiple source components and a background component, is fitted to the data, resulting in a complete spectral and spatial description of the field of view.
For the application to IACT data, the major challenge of such an approach is the construction of a robust background model.

In this work, we apply the 3D likelihood analysis to various test data recently made public by the \hess collaboration, using the open analysis frameworks \ct and \gp.
First, we show that, when using these tools in a classical analysis approach and comparing to the proprietary \hess analysis framework, virtually identical high-level analysis results, such as field-of-view maps and spectra, are obtained.
We then describe the construction of a generic background model from data of \hess observations, and demonstrate that a 3D likelihood analysis using this background model yields high-level analysis results that are highly compatible with those obtained from the classical analyses.
This validation of the 3D likelihood analysis approach on experimental data is an important step towards using this method for IACT data analysis, and in particular for the analysis of data from the upcoming Cherenkov Telescope Array (CTA).
}


\maketitle


\section{Introduction}
\label{sec:introduction}
With the advent of the Cherenkov Telescope Array \citep[CTA, see][]{CTA2018}, the field of ground-based \gam-ray astronomy is undergoing a major transformation.
This is not only because CTA, due to its greatly improved sensitivity with respect to current instruments, will offer a great discovery potential, but also because it will be operated as an open observatory.
Among many implications, this entails that an open software package will need to be provided for the high-level analysis of CTA data.
There are currently two open-source packages proposed as prototypes for this software package: \ct\footnote{\url{http://cta.irap.omp.eu/ctools}} \citep{knoedlseder2016} and \gp\footnote{\url{https://gammapy.org}} \citep{deil2017}.

In this paper, we apply both of these tools\footnote{We used the most recent versions of the packages available at the time of writing: \ct~\emph{1.6.0} and \gp~\emph{0.12}.} to the analysis of data from the High Energy Stereoscopic System (\hess), one of the current-generation arrays of imaging atmospheric Cherenkov telescopes (IACTs).
The motivation for this is twofold.
Firstly, we want to support the development of the tools.
While they have been validated in a CTA-internal data challenge based on simulated data, they have not been extensively tested on experimental data so far.
Secondly, both \ct and \gp\ -- despite still being in development -- already offer analysis techniques that are beyond the capabilities of the standard software packages used for high-level data analysis within the \hess Collaboration.
In particular, they allow us to perform a three-dimensional (3D), energy-resolved likelihood analysis, which is a major focus of this paper.

In a 3D likelihood analysis, the observed data are described by a combination of spectromorphological (i.e.\ three-dimensional) models, one for each relevant component in the observed field of view.
The models are fitted to the data via a likelihood formalism; the significance of specific components can be determined by means of likelihood ratio tests.
This kind of analysis is useful in cases where a `complicated' field of view, with multiple sources or large-scale diffuse emission, prevents standard ana\-lysis techniques from working well as they typically rely on a measurement of the residual cosmic-ray background within the observed field of view.
One of the major challenges in this approach is the development of an accurate model template for the cosmic-ray background, which strongly depends on the observation conditions.
In this paper, we attempt to construct such a model from archival \hess observations.

The concept of the 3D likelihood analysis is not new.
It was already applied in the analysis of data from the Energetic Gamma Ray Experiment Telescope (EGRET) \citep{Mattox1996}.
3D likelihood analysis is also routinely used in the analysis of data from the Fermi-LAT satellite \citep[see e.g.][]{FermiLAT2017}, for which changes and uncertainties in observation conditions are much less pronounced, and no strong residual cosmic-ray background is present.
Various forms of model-based analyses have also already been pioneered in IACT data analysis.
The construction of a background model template from cosmic ray-like events in the observed field of view is described in \citet{rowell2003} and \citet{fernandes2014}.
\citet{hessvelax2012} obtain an estimate for the cosmic-ray background by pairing each observation with one `Off' observation that is free of gamma-ray sources and has been taken under similar observation conditions -- this approach differs from the one presented in this paper in as much as the background is estimated from a single observation rather than many.
A morphological, energy-integrated likelihood analysis with multiple components is presented in \citet{hessarc2018,hgps2018}.
Finally, first fully spectromorphological likelihood analyses of \hess data have been carried out by \citet{mayer2014}, \citet{devin2018}, and \citet{ziegler2018}.

The outline of the paper is as follows:
Section~\ref{sec:hess} gives an overview about \hess and the data set utilised in this paper.
In Sect.~\ref{sec:standard}, we perform a validation of \ct and \gp.
This is achieved by performing several analyses of \hess data using standard IACT data analysis techniques, in particular standard techniques that treat the residual background of cosmic ray-induced air showers that are always present in the data.
We compare the results of the open-source tools with those obtained with the \hess analysis package (\hap), one of the proprietary \hess software packages used for data analysis.
In Sect.~\ref{sec:background}, we then introduce a novel method to construct a template model for the residual cosmic-ray background, a key prerequisite for the 3D likelihood analysis technique.
We then used \ct and \gp to apply this background model in a 3D likelihood analysis, as presented in Sect.~\ref{sec:template}, thereby also validating this analysis approach and exploring its capabilities.
Finally, we conclude the paper with Sect.~\ref{sec:conclusion}.

In several parts of this paper, we show results obtained with either one or both of the open-source science tools.
We stress that in every case, our intention is to demonstrate that both tools work well and yield results that are compatible with the \hess software package \hap, not to perform a comparison between the two.

We release the background model templates that we derived for the data analysed here as supplementary material to this paper, see Appendix~\ref{sec:appendix_bkg_model} for more information.
Furthermore, we make available the results of all spectral fits carried out for this paper in machine-readable format (see Appendix~\ref{sec:appendix_result_tables}).

\section{The High Energy Stereoscopic System and the \hess public test data release}
\label{sec:hess}
\hess \citep{hesscrab2006} is an array of five IACTs, located in the Khomas highland in Namibia ($23^\circ 16'18''$ S, $16^\circ 30'00''$ E), at an elevation of $1800\,\mathrm{m}$ above sea level.
In its first phase (`\hess Phase-I'), lasting from 2004 until 2012, the array consisted of four telescopes (CT1-4) with $107\,\mathrm{m}^2$ mirror area each, arranged in a square formation with $120\,\mathrm{m}$ side length.
In this configuration, \hess was able to detect \gam rays with energies from $\sim 200\,\mathrm{GeV}$ (for observations close to zenith) up to several tens of TeV.

The array was enhanced in 2012 by the addition of a fifth telescope (CT5) with mirror area $612\,\mathrm{m}^2$ in the centre of the array, thus reducing the energy threshold of the instrument to below $100\,\mathrm{GeV}$ \citep{holler2015}.
Data from this second phase of the experiment (`\hess Phase-II') are not analysed in this paper, however, all presented concepts can in principle also be applied to them.

\hess records data in time intervals of usually 28~minutes, called `observations' or `runs'.
The entire \hess Phase-I data set consists of $17\,712$ observations fulfilling basic quality criteria that check for hardware failures \citep[referred to as `detection' criteria, see][]{hesscrab2006}, amounting to $\approx 8\,050$~hours of observation time.
Of these, $15\,042$ observations ($6\,878$~hours) additionally pass a stricter set of quality criteria that ensures stable atmospheric conditions (`spectral' criteria).
About half of these observations are used in Sect.~\ref{sec:background} to construct a background model template.

We analyse data from the first \hess public test data release \citep{hesstestdata} in this paper.
Table~\ref{tab:data_sets} lists the data sets contained in this release.
Here, we use the data sets taken on the \crab, \pks (steady), \msh, and \rxj.
For each observation, the data consist of a list of recorded events with their reconstructed properties and instrument response functions (effective area, point spread function, and energy dispersion) specific to the observation.

\begin{table*}
  \caption{Data sets in the first \hess public test data release \citep{hesstestdata}.}
  \label{tab:data_sets}
  \centering
  \begin{tabular}{lcccccc}
    \hline
    Designation & Type & Extension & \# Runs & Observation time & Zenith angle & Publications\\
     & & & & (hours) & (deg) & \\\hline
    \crab & PWN & point-like & 4 & 1.9 & 45.4 -- 48.6 & (1,2) \\
    \pks (flare) & AGN & point-like & 15 & 7.0 & 7.2 -- 50.4 & (3,4) \\
    \pks (steady) & AGN & point-like & 6 & 2.8 & 22.8 -- 36.8 & (5,6,7) \\
    \msh & PWN & extended & 20 & 9.1 & 36.1 -- 40.2 & (8) \\
    \rxj & SNR & extended & 15 & 7.0 & 16.7 -- 26.3 & (9,10,11) \\
    Off runs & -- & -- & 45 & 20.7 & 2.5 -- 52.7 & -- \\\hline
  \end{tabular}
  \tablefoot{`Type' refers to the source type; PWN = pulsar wind nebula; AGN = active galactic nucleus; SNR = supernova remnant.
  `Extension' specifies whether or not the source can be spatially resolved using this data set.
  `Publications' lists (selected) \hess publications about the sources (usually based on larger data sets).
  The data sets `\pks (flare)' and `Off runs' are not used in this paper.}
  \tablebib{(1)~\citet{hesscrab2006}; (2)~\citet{holler2015}; (3)~\citet{hesspksflare2007}; (4)~\citet{hesspksflare2009};
  (5)~\citet{hesspks2005}; (6)~\citet{hesspks2010}; (7)~\citet{hesspks2017}; (8)~\citet{hessmsh1552_2005};
  (9)~\citet{hessrxj2004}; (10)~\citet{hessrxj2006}; (11)~\citet{hessrxj1713_2018}.}
\end{table*}

The test data release has been published specifically to support the development of open-source tools like \ct and \gp.
It contains data taken on both point-like and extended sources, making it a good choice of data set for this paper.
The data are available in FITS format\footnote{\url{https://fits.gsfc.nasa.gov}}, as specified by \citet{openspecs}, and can be directly processed with \ct and \gp.
We note that the data have been processed with an analysis configuration that is no longer state-of-the-art, both in terms of event reconstruction \citep[which is based on Hillas parameters;][]{Hillas1985} and \gam-hadron separation \citep[which uses the `mean scaled width' parameter;][]{hesscrab2006}.
This is not a problem, considering that the analysed sources are strong \gam-ray emitters and that the main purpose of this paper is the validation of the analysis tools.


\section{Validation of standard background estimation techniques}
\label{sec:standard}
In this section, we demonstrate the capability of the open-source analysis tools \ct and \gp to carry out analyses based on background estimation techniques that are traditionally used in ground-based \gam-ray astronomy.
We restrict ourselves to the two arguably most widely used techniques, namely the `ring background' and the `reflected background' algorithm \citep[for a detailed description of these algorithms, see][]{berge2007}.
Both techniques apply aperture photometry, that is, they extract the flux of \gam rays from a source by determining the number of registered events in a region of interest, called `on region', and comparing this to an (appropriately scaled) estimate of the residual cosmic-ray background, obtained from one or multiple `off region(s)' within the observed field of view.

Furthermore, we validate the results obtained with the open-source tools by comparing them with results obtained with the \hess-internal analysis software package \hap.
In all cases, we find the results to be virtually identical.

We present results obtained with the ring background method in Sect.~\ref{sec:ring}, those obtained with the reflected background method in Sect.~\ref{sec:reflected}.
In some cases, we only show results obtained with one of the open-source tools, or for a selection of the available data sets, implying that we obtain the same level of agreement with the other tool or the remaining data sets, respectively.
Table~\ref{tab:analysis_settings} lists the settings used in the analyses (common for all tools).

\begin{table*}
  \caption{Settings for analyses applying standard background estimation techniques.}
  \label{tab:analysis_settings}
  \centering
  \begin{tabular}{lccccc}
    \hline
    Source & R.A. & Dec. & $\theta^2$ & $r_\mathrm{min}$ & $r_\mathrm{max}$ \\
     & (deg) & (deg) & (deg$^2$) & (deg) & (deg) \\\hline
    \crab & 83.63 & 22.01 & 0.0125 & 0.5 & 0.7 \\
    \pks & 329.72 & $-30.23$ & 0.0125 & 0.5 & 0.7 \\
    \msh & 228.53 & $-59.16$ & 0.09 & 0.5 & 0.7 \\
    \rxj & 258.39 & $-39.76$ & 0.36 & 0.6 & 0.8 \\\hline
  \end{tabular}
  \tablefoot{`R.A.' and `Dec.' give the source position in equatorial coordinates (J2000) as used in the analysis.
  $\theta^2$ denotes the squared radius of the on region containing the source.
  $r_\mathrm{min}$ and $r_\mathrm{max}$ are the inner and outer radius used in the ring background algorithm.}
\end{table*}

\subsection{Ring background method}
\label{sec:ring}
The ring background method is typically used to visualise the excess of \gam rays attributed to a source, either in the form of a one-dimensional `$\theta^2$-plot'\footnote{The angular distance between the reconstructed direction of an event and the source is usually denoted with $\theta$.} or of two-dimensional sky maps.
In both cases, the algorithm starts from a binned map of the events registered in the observation.
For each pixel, it then determines an estimate of the residual cosmic-ray background for that pixel by summing up the events in all pixels contained in a ring around the pixel with inner radius $r_\mathrm{min}$ and outer radius $r_\mathrm{max}$ (cf. Table~\ref{tab:analysis_settings}).
Pixels around known \gam-ray emitters need to be excluded in this process.

The background estimate must be corrected for the different exposure of the pixels in the ring with respect to the pixel under consideration.
Since the acceptance (i.e.\ the probability of detecting an event) of the experiment varies across the field of view, a model of the acceptance is required to apply this correction.
For the sake of better comparability, we chose to employ the model utilised in the \hap analysis within the analyses carried out with \ct and \gp.
The model is based on archival \hess data, similar to the model introduced in Sect.~\ref{sec:background}, but less detailed (e.g.\ a radial symmetry is assumed).
We verified that we obtain compatible results when using the background model developed in this paper instead.

A potential excess of \gam-ray events can then be determined by subtracting the background estimate from the map of registered events.
Similarly, the significance of the excess can be computed.

\subsubsection{\texorpdfstring{$\theta^2$}{Theta2} distributions}
\begin{figure}
  \centering
  \includegraphics{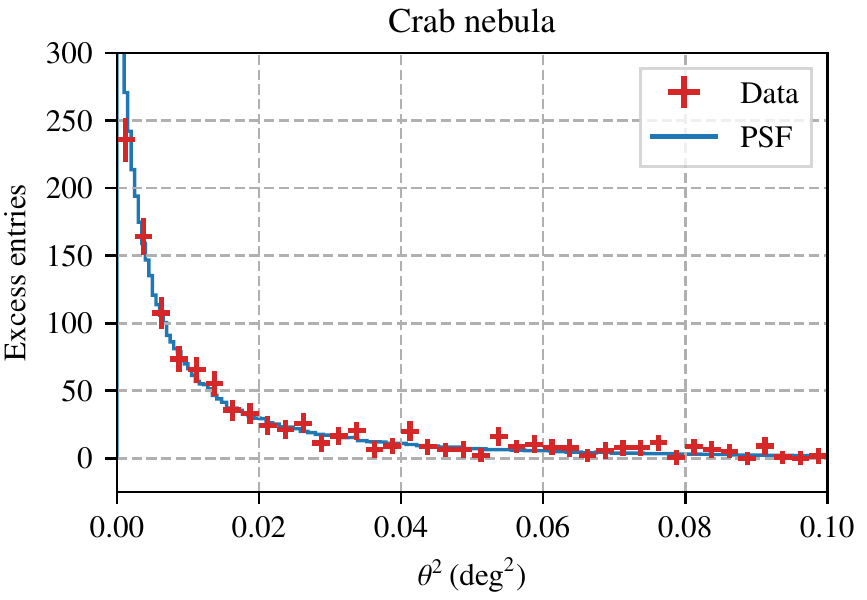}
  \caption{Distribution of squared angular distance $\theta$ between reconstructed event directions and source position for the \crab, obtained with \gp.
  The blue line shows the shape of the point spread function of the instrument for this data set.}
  \label{fig:theta2_crab}
\end{figure}

Especially for point-like sources, a $\theta^2$ distribution is often used to display the excess of \gam rays from the direction of the source.
In Fig.~\ref{fig:theta2_crab} we show such a distribution for the \crab data set, here obtained with \gp.
The shape of the point spread function (PSF) for this data set, averaged over all energies and assuming the best-fit energy spectrum for this source (see Sect.~\ref{sec:reflected}), is illustrated as well.
As expected for a point-like source\footnote{Even though the \crab has been demonstrated to be extended in very-high-energy \gam rays recently \citep{holler2017}, it can be considered point-like for the data set and analysis configuration used here.}, the distribution follows the shape of the PSF, demonstrating that this instrument response function is processed correctly by \gp.

\subsubsection{Sky maps}
The \gam-ray excess may also be visualised in the form of sky maps, in particular in the case of spatially extended sources.
Different quantities can be plotted; here we focus on the common case of sky maps denoting the significance of an excess.
As is custom, we smoothed statistical fluctuations by convolving the map of registered events with a top-hat kernel of $0.1\,\mathrm{deg}$ radius \citep[see e.g.][]{hgps2018} and computed the significance following \citet{Li1983}.

\begin{figure}
  \centering
  \includegraphics{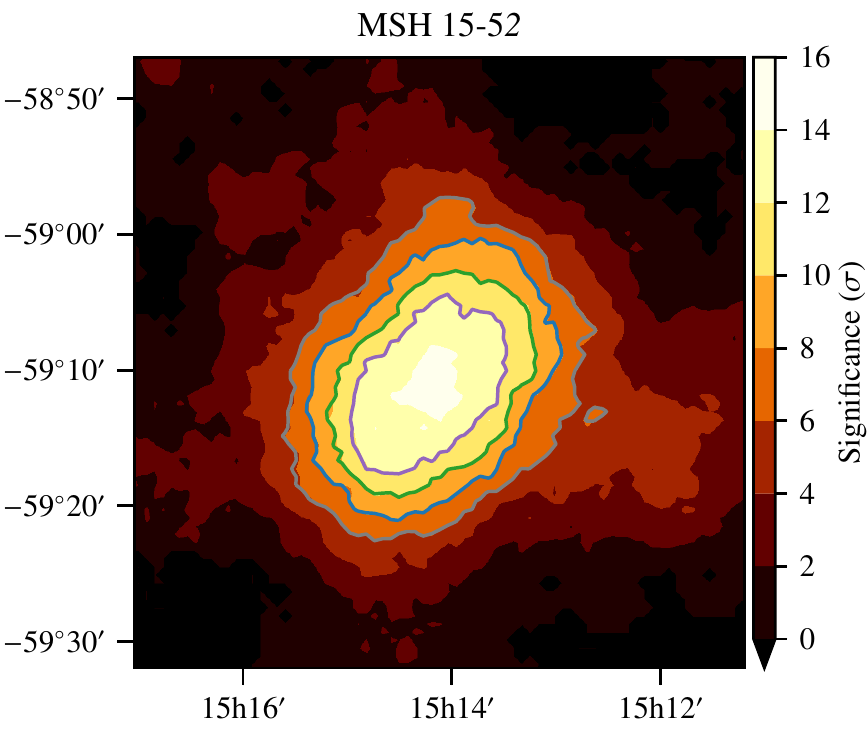}
  \caption{Significance map for pulsar wind nebula \msh, in equatorial coordinates (J2000).
  The map in the background has been derived with \gp, the coloured lines display 6, 8, 10, and 12 $\sigma$ confidence level contours for the corresponding map derived with \hap.}
  \label{fig:map_msh1552}
\end{figure}

\begin{figure}
  \centering
  \includegraphics{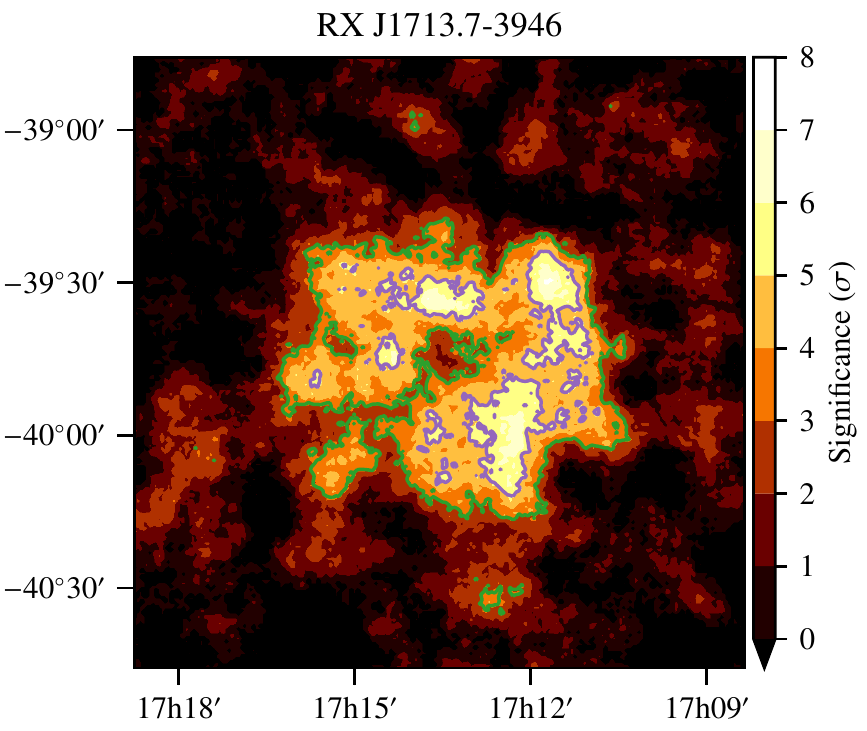}
  \caption{Significance map for supernova remnant \rxj, in equatorial coordinates (J2000).
  The map in the background has been derived with \ct, the coloured lines display 3 and 5 $\sigma$ confidence level contours for the corresponding map derived with \hap.}
  \label{fig:map_rxj1713}
\end{figure}

We show maps for the two extended sources that are part of the \hess public test release data set, \msh and \rxj, in Figs.~\ref{fig:map_msh1552} and~\ref{fig:map_rxj1713}, respectively.
In both cases, we plot contour lines of the map derived with the \hap software on top of the map derived with one of the open-source tools, observing an extremely good agreement.

\begin{figure}
  \centering
  \includegraphics{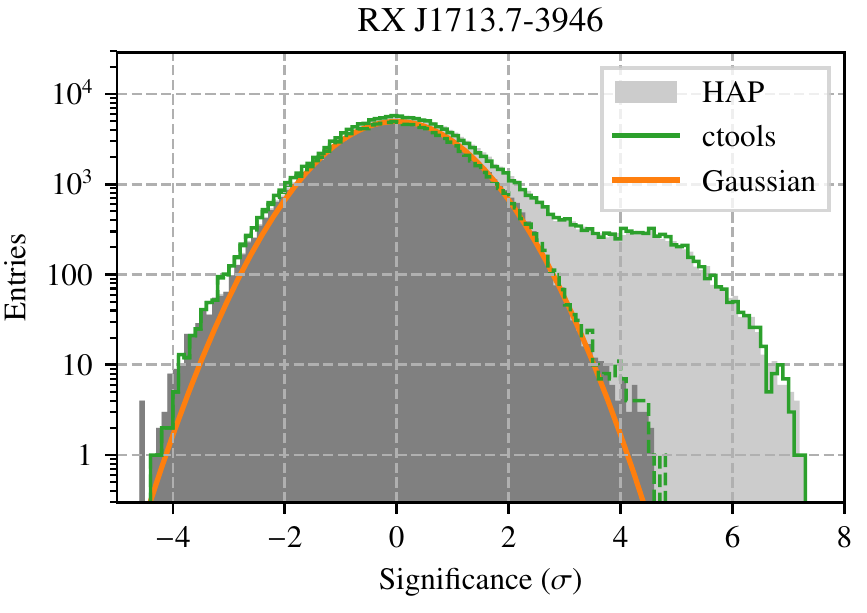}
  \caption{Significance distribution for \rxj.
  The histograms are entry distributions of the significance map shown in Fig.~\ref{fig:map_rxj1713}, obtained with \hap (grey filled histograms) and \ct (green histograms).
  The distributions are shown for pixels outside source regions (dashed green, dark grey) and for all pixels (solid green, light grey).
  The orange line shows a Gaussian distribution with zero mean and unity width.}
  \label{fig:sign_rxj1713}
\end{figure}

The quality of the description of the cosmic-ray background in the field of view can be judged by inspecting the entry distribution of a significance map.
Figure~\ref{fig:sign_rxj1713} displays the distributions for the map shown in Fig.~\ref{fig:map_rxj1713}.
For a perfectly modelled background, the significance distribution of pixels outside source regions approaches that of a Gaussian distribution (shown by an orange line, for comparison).
As for the maps themselves, we observe a very good agreement between the tools.

\subsection{Reflected background method}
\label{sec:reflected}
The reflected background method is usually employed to determine the flux of \gam rays from a given source, in other words, to extract its spectrum.
It requires the observations to be carried out in so-called `wobble mode', meaning that the source is observed under a small offset with respect to the pointing direction of the telescopes.
This allows the definition of off regions that are `reflected' about the pointing direction, meaning that they have the same offset to the pointing direction as the on region encompassing the source.
The acceptance can then be assumed to be approximately the same in all regions, leading to reduced systematic uncertainties in the background determination.
The extraction of the spectrum then proceeds by determining the excess of \gam-ray events in the on region with respect to the off regions and performing a forward-folding likelihood fit, utilising the instrument response functions (here, the effective area and the energy dispersion matrix).
For this paper, we always assumed that the spectrum has the form of a power law,
\begin{linenomath*}\begin{equation}\label{eq:powerlaw}
  \frac{\mathrm{d}N}{\mathrm{d}E} = \phi\times \left(\frac{E}{E_0}\right)^{-\Gamma}\,,
\end{equation}\end{linenomath*}
where $\phi$ and $\Gamma$ are free parameters and $E_0$ is a normalisation energy that we chose such that the correlation between $\phi$ and $\Gamma$ is minimised.
We furthermore computed spectral points with a fixed binning of eight flux points per decade of energy, assuming that the spectral index in each bin is equal to that of the fitted power law.

In order to ensure an accurate reconstruction of the arrival direction of the primary particle, a minimum signal strength is usually required for each telescope (for the analysis configuration used for the data we analysed here, this signal threshold is set at 80 photo-electrons for each camera image).
This signal threshold per telescope translates into a minimum energy of the primary \gam ray.
\gam rays around and below that energy can only be detected (i.e.\ pass the signal threshold) if an upward fluctuation of the signal occurs.
This typically leads to a bias in the reconstructed energy of \gam rays at these energies.
While this can be corrected for in the extraction of the energy spectrum, it is usually a good measure to define an analysis energy threshold that ensures that the bias is not too large.
Here, we required that the bias of the energy reconstruction of \gam-ray events incident under an offset angle corresponding to that of the location of the observed source does not exceed 10\%.

We determined the positions of off regions with each of the tools separately, requiring that no regions are placed at the location of known \gam-ray sources\footnote{
Due to the presence of other sources, it is not possible to find off regions for four of the observations in the \rxj data set; we excluded these observations from the spectral analysis.
Furthermore, the algorithm implemented in \ct did not find off regions for an even larger number of observations.
We therefore used the off regions determined with \gp for the extraction of the spectrum for \rxj with \ct.}.
We also computed analysis energy thresholds as described above with \hap and \gp and applied them in the extraction of the energy spectrum with these tools, respectively.
Since \ct in its current version does not provide the possibility to compute energy thresholds based on the energy reconstruction bias, we used the thresholds determined with \gp also for the spectrum extraction with \ct.
We furthermore note that \ct, unless provided with instrument response functions specifically prepared for point sources, always applies a correction to the effective area based on an assumed leakage of \gam-ray events outside the defined source region.
This is not desired in the reflected background analysis of the two extended sources studied here (\msh and \rxj), since we chose on regions that encompass the sources by a large enough margin such that the leakage is negligible; this is a standard procedure in \hess data analysis.
For the sake of better comparability, we therefore decided to actively disable this correction by modifying the \ct source code.

\begin{figure}
  \centering
  \includegraphics{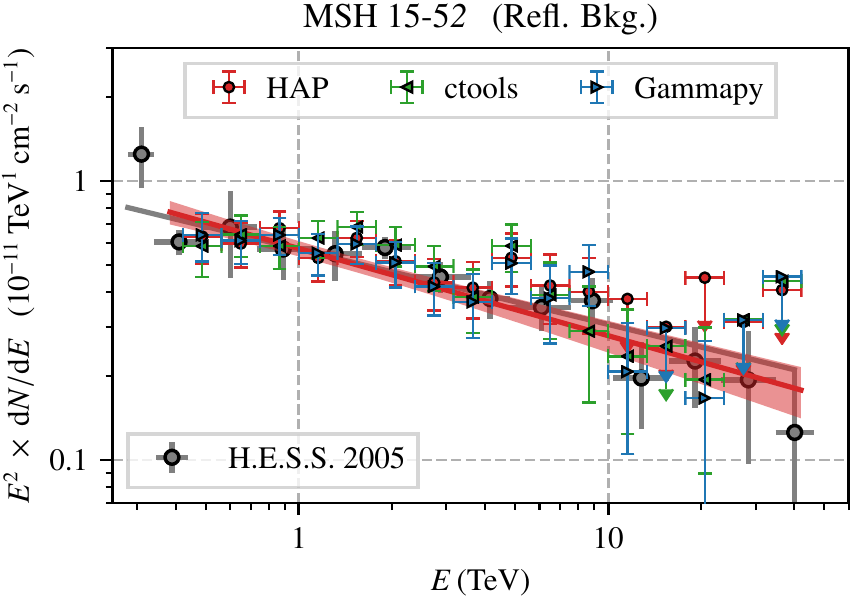}
  \caption{Comparison of spectra for \msh.
  The spectra shown in red, green, and blue were derived with the reflected background method with \hap, \ct, and \gp, respectively.
  For the \hap analysis, we show the result of the power-law fit in addition.
  We compute upper limits (95\% confidence level) for flux points with a statistical significance of less than two standard deviations.
  The published spectrum is taken from \citet{hessmsh1552_2005}.}
  \label{fig:spectrum_standard_msh1552}
\end{figure}

We show energy spectra extracted with the reflected background method for the \msh and \rxj data sets in Figs.~\ref{fig:spectrum_standard_msh1552} and~\ref{fig:spectrum_standard_rxj1713}, respectively.
Spectra derived for other sources can be found in Appendix~\ref{sec:appendix_reflected}.
Finally, we compare the best-fit parameter values for the normalisation and spectral index of the power law model obtained with the different tools for all sources in Fig.~\ref{fig:pl_par_dev}.
Table~\ref{tab:fitresults} in Appendix~\ref{sec:appendix_fit_results} furthermore lists the best-fit parameter values for all analyses.
In general, we observe an excellent agreement between the three different tools for all spectra, both concerning the power-law fits and the spectral flux points.
The remaining differences give an indication of the systematic uncertainties associated with the implementation of the analysis technique.

\begin{figure}
  \centering
  \includegraphics{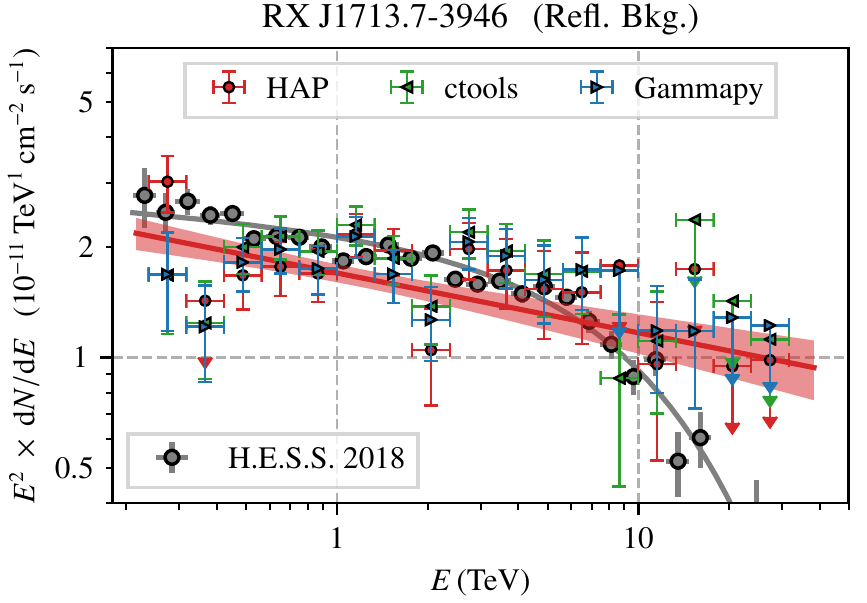}
  \caption{Comparison of spectra for \rxj.
  The spectra shown in red, green, and blue were derived with the reflected background method with \hap, \ct, and \gp, respectively.
  For the \hap analysis, we show the result of the power-law fit in addition.
  We compute upper limits (95\% confidence level) for flux points with a statistical significance of less than two standard deviations.
  The published spectrum is taken from \citet{hessrxj1713_2018}; it uses a power law with exponential cut-off as spectral model.}
  \label{fig:spectrum_standard_rxj1713}
\end{figure}

\begin{figure}
  \centering
  \subfigure[Flux normalisation $\phi$.]{
    \includegraphics{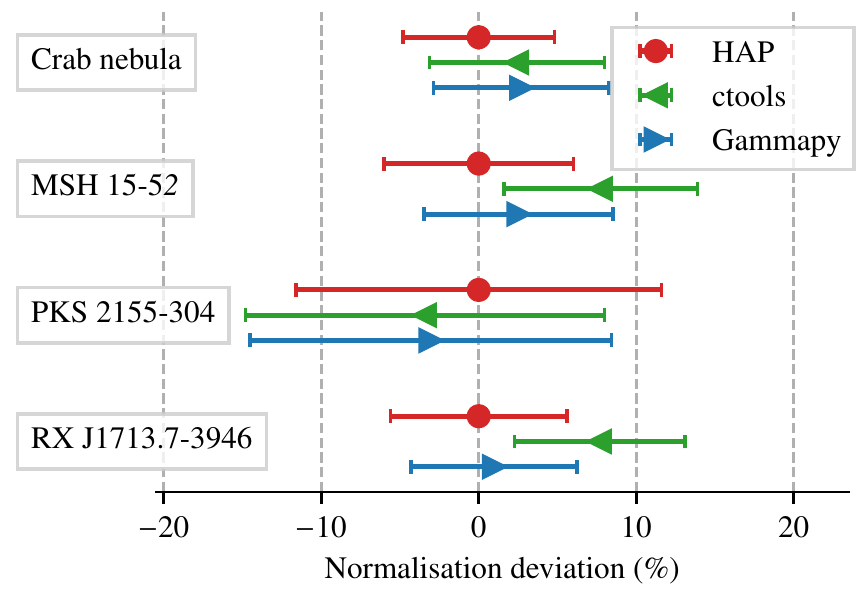}
    \label{fig:pl_norm_dev}
  }\\
  \subfigure[Spectral index $\Gamma$.]{
    \includegraphics{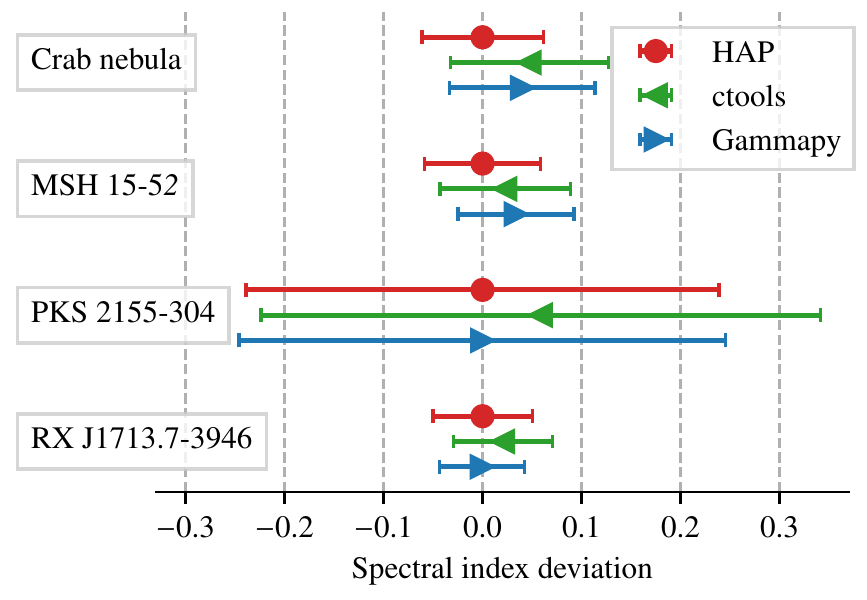}
    \label{fig:pl_index_dev}
  }
  \caption{Comparison of spectral fit parameters for all sources.
  Displayed are the results obtained using the reflected background method with \hap, \ct, and \gp.
  We plot the deviation w.r.t.\ the results obtained with our standard tool \hap.
  The error bars denote statistical uncertainties only (68\% confidence level).}
  \label{fig:pl_par_dev}
\end{figure}

It is noteworthy that the spectrum extracted for \rxj (cf.\ Fig.~\ref{fig:spectrum_standard_rxj1713}) does not agree very well with the spectrum published in \citet{hessrxj1713_2018} at energies below $\sim 0.45\,\mathrm{TeV}$.
We attribute this discrepancy to the fact that the reflected background method is not well suited for sources with an extent as large as \rxj\footnote{A more appropriate method to determine the residual cosmic-ray background has been used in \citet{hessrxj1713_2018}.}.
Indeed, we were able to find only a single off region for almost all of the observations in this data set, which makes the analysis susceptible to both statistical and systematic uncertainties that are normally reduced by averaging over multiple off regions.
However, our main purpose in this section being the validation of the open-source tools, we note that the tools actually agree well also in this region and do not investigate the discrepancy to the published spectrum further.


\section{Development of a background model template}
\label{sec:background}
In this section we introduce a procedure to construct a model for the residual cosmic-ray background in arbitrary field of views from archival \hess observations.
We note that the archival data are proprietary to the \hess Collaboration and not publicly available.
The procedure is inspired by the work of \citet{mayer2014}, but has been considerably advanced in the course of this work.
The constructed model yields the expected background rate as a function of two field-of-view coordinates and the reconstructed energy of the primary particle, which means it is three-dimensional.
We present the construction procedure in Sect.~\ref{sec:construct}, before we characterise and validate the background model in terms of its spatial and spectral properties in Sect.~\ref{sec:characterise_validate}.
We make available as supplementary material to this paper the background model for the observations that are contained in the first \hess public test data release (see Appendix~\ref{sec:appendix_bkg_model} for more information).

\subsection{Construction}
\label{sec:construct}
We begin with the selection of archival \hess observations that are used to construct the model.
Here we considered only observations taken during the first phase of \hess, without the large CT5 telescope.
Since we aim to model the residual cosmic-ray background only, we excluded observations taken in the direction of the Galactic plane ($|l|<60^\circ, |b|<5^\circ$), since we expect contamination from diffuse \gam-ray emission there.
We furthermore imposed `spectral' observation quality criteria, rejecting observations that have been taken in the presence of hardware failures or under bad atmospheric conditions \citep[see][for more information]{hesscrab2006}.
Finally, we used only observations in which all four small telescopes have participated in data taking.
This selection yields $7\,063$ observations with a total observation time of $\approx 3\,240$~hours, taken between January 21, 2004 and May 15, 2013.

The background rate depends on various observational parameters, requiring us to take these into account and construct a tailored background model for each observation.
The construction procedure consists of two steps.
We first created an initial model that takes the principal dependencies of the background rate into account.
This initial model was subsequently refined in an iterative procedure, thus correcting for less pronounced effects.

\subsubsection{Initial model}
\label{sec:init_model}
For the initial model, we first considered the pointing direction of the telescopes in the horizontal (i.e.\ local) coordinate system.
This is motivated by the strong dependence of the background rate on the zenith angle of the observation (cf.\ the characterisation of the model in Sect.~\ref{sec:characterise}).
The dependence on the azimuth angle (due to the Earth's magnetic field), albeit less strong, could easily be incorporated into the model at this stage as well.

To take these effects into account, we grouped the observations in bins of the average zenith angle ($\vartheta$) and azimuth angle ($\phi$) of their pointing direction and constructed an initial model for each of these bins.
As the background rate does not vary strongly with azimuth angle, it is sufficient to use only two bins for this parameter ($-90^\circ<\phi<90^\circ$ and $90^\circ<\phi<270^\circ$).
The dependence of the background rate on the zenith angle is much stronger, in particular as the zenith angle increases.
Here we used eight bins for this parameter, as listed in Table~\ref{tab:bg_bin_data}.
The table furthermore lists the resulting number of observations (and corresponding observation time) available in each of the bins.

\begin{table}
  \caption{Background model binning and statistical information.}
  \label{tab:bg_bin_data}
  \centering
  \begin{tabular}{ccccc}
    \hline
    & \multicolumn{2}{c}{$-90^\circ<\phi<90^\circ$} & \multicolumn{2}{c}{$90^\circ<\phi<270^\circ$}\\\hline
    $\vartheta$ & $N_\mathrm{obs}$ & $t_\mathrm{live}$ & $N_\mathrm{obs}$ & $t_\mathrm{live}$ \\
    (deg) & & (hours) & & (hours) \\\hline
    $0-10$ & 99 & 44.5 & 660 & 301.8 \\
    $10-20$ & 392 & 177.8 & 994 & 455.6 \\
    $20-30$ & 650 & 297.0 & 1378 & 632.2 \\
    $30-40$ & 444 & 201.4 & 790 & 367.6 \\
    $40-45$ & 300 & 135.8 & 242 & 110.9 \\
    $45-50$ & 306 & 140.0 & 448 & 204.9 \\
    $50-55$ & 150 & 68.4 & 124 & 57.3 \\
    $55-60$ & 61 & 28.2 & 25 & 12.2 \\\hline
  \end{tabular}
  \tablefoot{The number of observations $N_\mathrm{obs}$ used to construct the initial background model for each bin of zenith angle $\vartheta$ and azimuth angle $\phi$.
  The corresponding live time $t_\mathrm{live}$ is listed as well.}
\end{table}

Further parameters that affect the background rate and could thus be considered in the construction of the model include for example the date of the observation (accounting for efficiency loss over time) or the atmospheric conditions during the observation.
However, a too fine separation of observations into bins can lead to an insufficient number of observations per bin, resulting in too large statistical uncertainties.
We therefore chose not to add further dimensions to the initial model but rather to incorporate a correction for these effects in a later step (see Sect.~\ref{sec:refine}).

\begin{figure*}
  \sidecaption
  \includegraphics{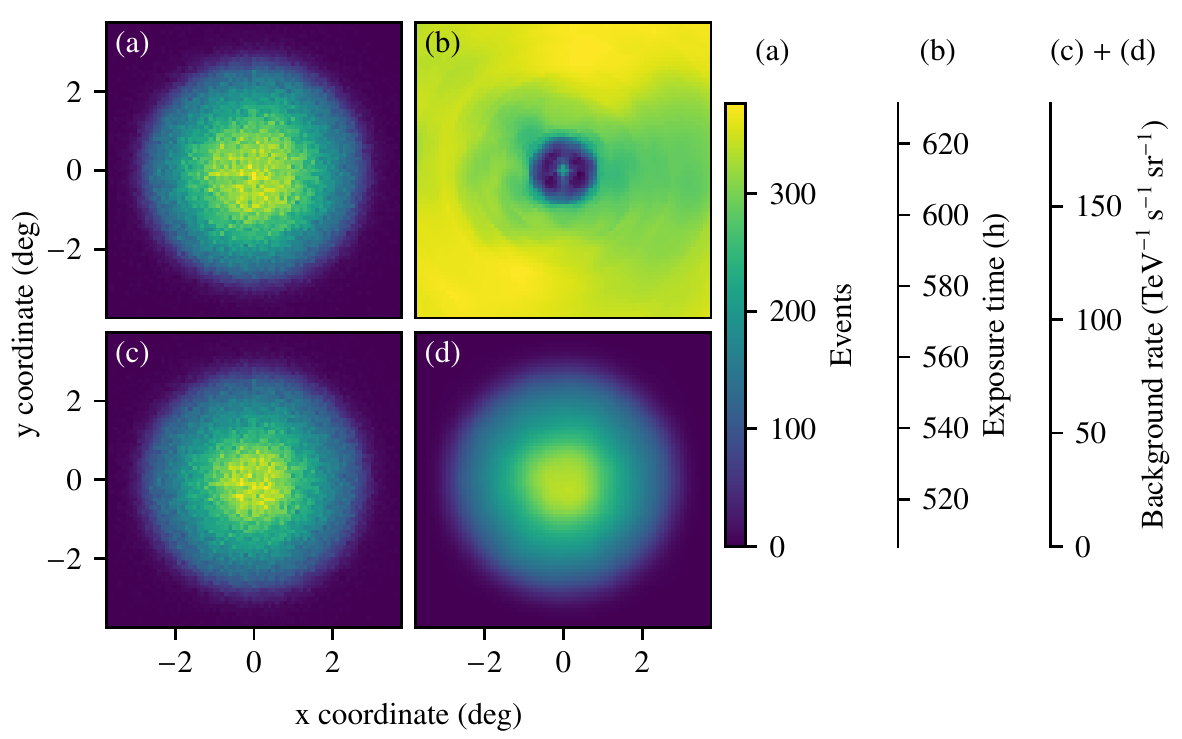}
  \caption{Illustration of background model construction.
  We exemplary show the model for the bin with azimuth angle $90^\circ<\phi<270^\circ$ and zenith angle $20^\circ<\vartheta<30^\circ$, for energies between 0.8 and 1.1~TeV.
  (a) Sum of the number of registered events in each observation, excluding events with a reconstructed direction close to known \gam-ray sources.
  (b) Effective exposure time of each spatial pixel.
  The effective exposure time is the summed observation time of all observations, corrected for the exclusion of events from regions around known \gam-ray sources.
  (c) Averaged background rate, given by the number of registered events divided by effective exposure time, energy interval and solid angle.
  (d) Averaged background rate after the application of a spline-based smoothing algorithm.
  All vertical axis labels refer to the same colour bar.}
  \label{fig:bg_construct}
\end{figure*}

We constructed the model in a field-of-view coordinate system that is aligned with the horizontal coordinate system, but rotated such that the pointing position of the corresponding observation lies at the equator at coordinates $(l=0,b=0)$.
In this system, the longitude axis points in the direction of decreasing azimuth angles, and the latitude axis points in the direction of increasing altitude angles (or decreasing zenith angles, respectively).
This allowed us to compute an average background rate from all observations in each bin of $\vartheta$ and $\phi$, even if their pointing positions in equatorial coordinates are different.
We obtained an averaged rate in a square grid with a side length of $7.5^\circ$ and spatial bins of $0.1^\circ$ size.
The energy axis is divided into 20 logarithmically spaced bins between 100~GeV and 100~TeV.
In the computation, we discarded events from directions close to known \gam-ray sources, applying a corresponding correction to the exposure time per spatial bin.
The construction procedure is illustrated in Fig.~\ref{fig:bg_construct}~(a)-(c).

In the next step, we transformed the model into a field-of-view coordinate system that is aligned with the equatorial coordinate system.
Similar to the altitude-azimuth-aligned system, this system is centred on the pointing position, but rotated such that the longitude and latitude axes are aligned with the right-ascension and declination coordinate, respectively.
\ct requires any background model to be specified in this system, and \gp accepts models defined in this system as well.
Finally, we smoothed the initial model using a two-dimensional cubic spline function to remove statistical fluctuations that are inevitably introduced by the procedure described above (see Fig.~\ref{fig:bg_construct}~(d)\footnote{We note that, for illustrative purposes, the figure displays the smoothed model still in the altitude-azimuth-aligned field-of-view coordinate system, whereas we actually apply the smoothing algorithm \textit{after} transforming into the R.A./Dec.-aligned field-of-view coordinate system.}).

\subsubsection{Refined model}
\label{sec:refine}
In order to account for further observation-specific parameters in the background model construction, we need to assign an initial model to each individual observation.
Here, we start by simply grouping all observations in the same azimuth and zenith angle bins as used in the background model construction (cf.\ Table~\ref{tab:bg_bin_data}), and selecting for each observation the initial model that was constructed from observations falling into the same bin.

To assess the quality of the background model, and to apply corrections, we performed a likelihood fit of the model for each observation to the observed data, masking regions that contain known \gam-ray sources.
To avoid being dominated by bins at low energies, where statistics are large, we performed a fit in each energy bin separately, fitting the model normalisation in each bin.
We did not perform a fit for energy bins that are below the energy threshold computed for the observation.
We then obtained a single normalisation value for each observation by averaging over all energy bins.

\begin{figure}
  \centering
  \includegraphics{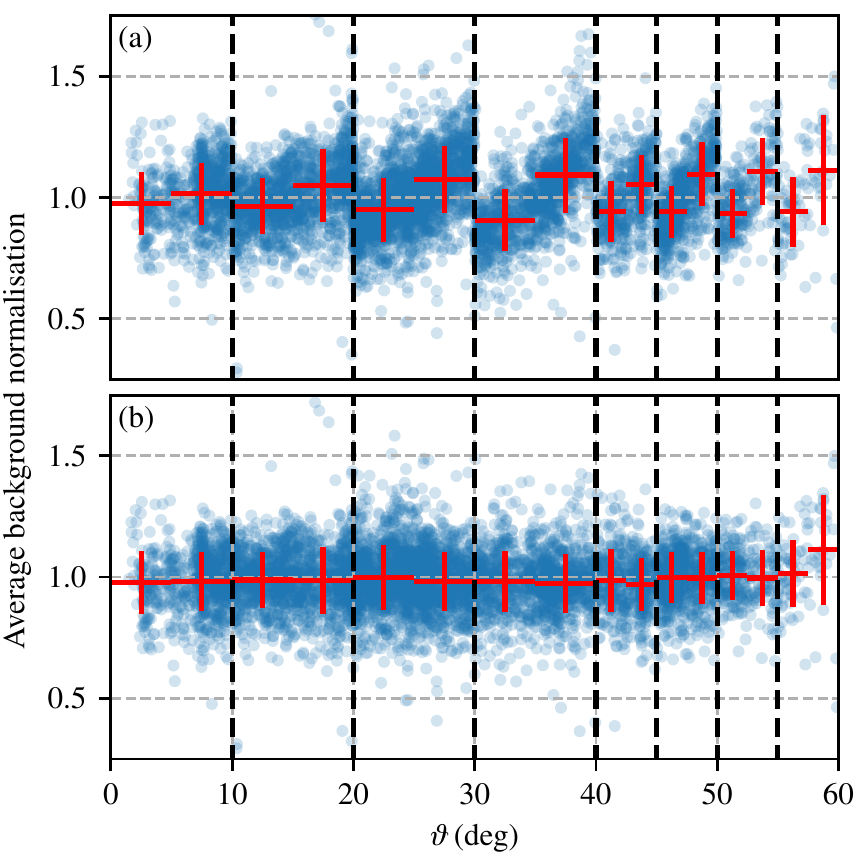}
  \caption{Average fitted background model normalisation as a function of zenith angle $\vartheta$:
  (a) without zenith-angle interpolation; (b) with zenith-angle interpolation.
  The normalisation averaged over energy bins is shown for all observations used in the model construction.
  Boundaries between zenith angle bins as used in the construction of the model are marked by the black dashed lines.
  The blue markers show the individual observations, whereas the red data points denote the mean and standard deviation in bins of $\vartheta$.}
  \label{fig:bg_norm_zenith}
\end{figure}

Figure~\ref{fig:bg_norm_zenith}~(a) shows the average background model normalisation obtained in this way for all observations that have been used to construct the model itself, as a function of the zenith angle $\vartheta$ of the observation.
It is evident that the procedure of selecting an initial model simply based on the zenith angle bin leads to jumps in the average fitted normalisation at the boundaries between bins.
We therefore proceeded to performing a linear interpolation of the predicted background rate between adjacent zenith angle bins to assign a model to each observation.
Re-performing the likelihood fit for all observations with this model, we observe that the average fitted background normalisation no longer depends on the zenith angle, as shown in Fig.~\ref{fig:bg_norm_zenith}~(b).

\begin{figure}
  \centering
  \includegraphics{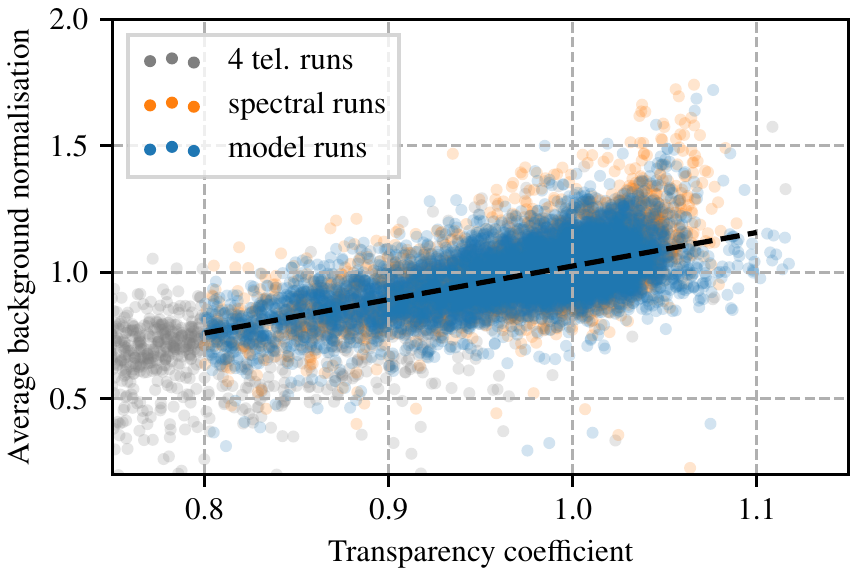}
  \caption{Average fitted background model normalisation as a function of transparency coefficient.
  The grey points denote all observations taken with four active telescopes.
  Observations fulfilling `spectral' quality criteria in addition are shown in orange.
  The observations that were used to construct the background model are depicted in blue.
  The black, dashed line shows the fit function that is used to correct for the atmospheric transparency.}
  \label{fig:tc_corr}
\end{figure}

In Fig.~\ref{fig:tc_corr}, we show the average fitted background normalisation as a function of the so-called transparency coefficient.
This coefficient is computed based on the trigger rate of the telescopes and describes the optical transparency of the atmosphere (in arbitrary units), with larger values implying a more transparent atmosphere \citep[for more details see][]{hahn2014}.
Unsurprisingly, the fitted background normalisation is correlated with the transparency coefficient (Spearman correlation coefficient~0.57), reflecting the fact that a decreased absorption of Cherenkov photons in the atmosphere leads to an increased rate of triggered background events.
The dependence can be fitted well with a linear function, as indicated in the figure.
Aiming for a more accurate description of the background rate, we used this function as a correction in the construction of the initial model as well as in the assignment of the initial model to individual observations.
After this correction, the fitted background normalisation is no longer correlated with the transparency coefficient.

As a last step, we applied a correction factor for the so-called `optical phase' of each observation (see Table~\ref{tab:bg_mu_correction}).
Optical phases are specific periods in time that have been defined in order to account for the varying optical efficiency of the telescopes, for example due to mirror degradation\footnote{The series of comparatively short periods in 2010 and 2011 are motivated by an exchange of the mirrors of the \hess telescopes performed during that time.}.
One set of Monte Carlo simulations for the generation of instrument response functions has been generated by the \hess Collaboration for each optical phase.
We observe a slight bias of the fitted background normalisations with respect to the expected value of~1 for some of the phases, which we attribute to imperfections in the Monte Carlo simulations.
Using the average fitted background normalisation for each phase (listed in Table~\ref{tab:bg_mu_correction}) as correction factor, we were able to eliminate this bias.

\begin{table}
  \caption{Correction factors for different `optical phases'.}
  \label{tab:bg_mu_correction}
  \centering
  \begin{tabular}{ccc}
    \hline
    Phase & Start & Correction factor\\\hline
    1 & 2004-01-21 & 1.145\\
    1b & 2004-05-26 & 0.994\\
    1c & 2007-07-03 & 0.948\\
    1c1 & 2010-04-27 & 0.960\\
    1c2 & 2010-10-17 & 1.047\\
    1c3 & 2011-04-14 & 1.082\\
    1d & 2011-11-12 & 1.004\\\hline
  \end{tabular}
  \tablefoot{Correction factor applied for each optical phase.
  The start date of each phase is given, with each phase lasting up to the start of the subsequent one.
  See main text for details.}
\end{table}

\begin{figure}
  \centering
  \includegraphics{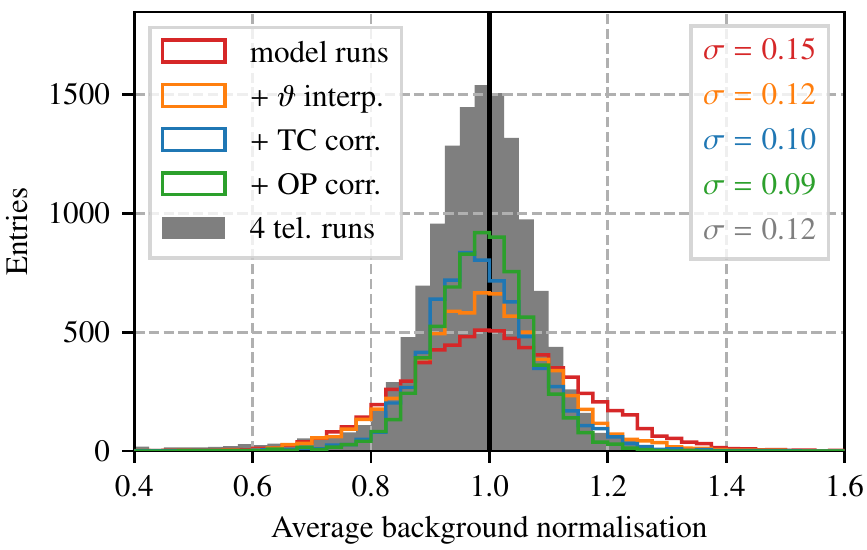}
  \caption{Distribution of fitted background normalisations.
  The histograms display the fitted normalisation values for all energy bins, i.e.\ the normalisations are not averaged for each observation here.
  The grey histogram shows the distribution for the final model and all 4-telescope observations.
  The coloured histograms display the distribution for the observations that have been used to construct the model, without corrections (red), with interpolation between zenith angle ($\vartheta$) bins (orange), with a correction for atmospheric transparency (blue), and with a correction for optical phases (green).
  We give the standard deviation of the distributions in the upper right corner of the plot.}
  \label{fig:fit_norms}
\end{figure}

We illustrate the improvement in accuracy achieved by the refinement procedure in Fig.~\ref{fig:fit_norms}.
While the distribution of fitted background normalisations has a standard deviation of 15\% for the initial model (considering only observations used in the model construction), we obtain a width of 9\% for the refined model.
Including also observations that were not used to construct the model, the standard deviation increases to 12\%.
This is expected, considering that observations not fulfilling spectral quality criteria or taken in the direction of the Galactic plane cannot be perfectly described by the background model.

Finally, we note that the background model often fails to describe the data well close to the energy threshold of the instrument, where the variation of the background rate with energy is large and strongly dependent on the specific observation conditions.
Therefore, it is usually necessary to apply an analysis energy threshold, thus restricting the analysis to energies where the background model describes the cosmic-ray background well.

\subsection{Characterisation and validation}
\label{sec:characterise_validate}

\subsubsection{Characterisation}
\label{sec:characterise}
\begin{figure}
  \includegraphics{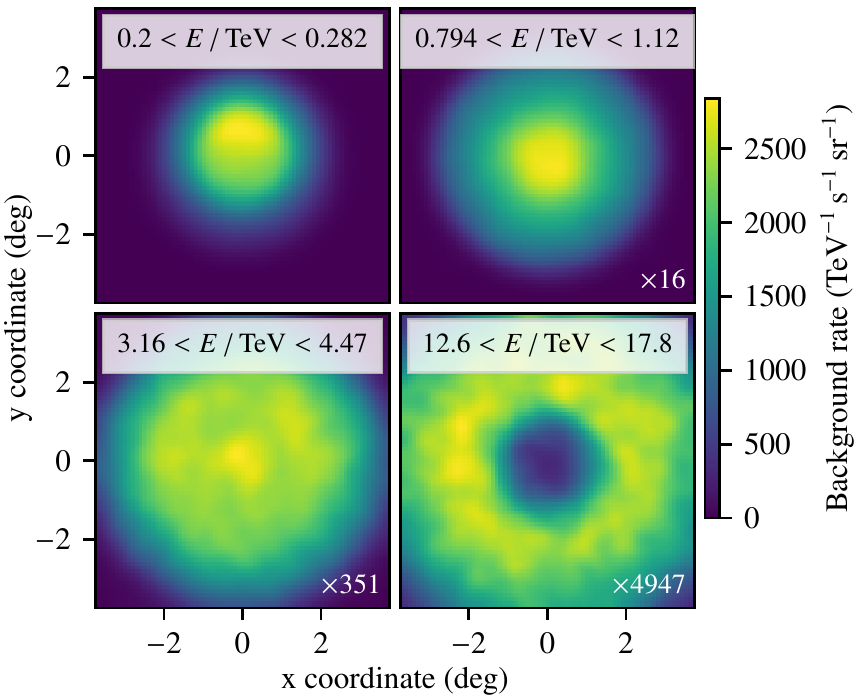}
  \caption{Background model visualisation.
  We show the background model in the field-of-view coordinate system, in four different energy bins.
  Here, the model for the bin with azimuth angle $90^\circ<\phi<270^\circ$ and zenith angle $20^\circ<\vartheta<30^\circ$ is displayed.
  The rate in all energy bins but the first has been multiplied by the factor indicated in the figure, to allow for a common colour scale.}
  \label{fig:bg_energy_maps}
\end{figure}

Figure~\ref{fig:bg_energy_maps} shows a visualisation of the spatial shape of the final model in different energy bins.
We show the model in the altitude-azimuth aligned field-of-view coordinate system here, that is, before the assignment to a specific observation.
The background rate is clearly asymmetric with respect to the y-coordinate in the first energy bin shown.
This reflects the altitude-angle (or zenith-angle) dependence of the background rate, since the y-coordinate is aligned with the altitude coordinate in the chosen coordinate system.
Around $1\,\mathrm{TeV}$, the shape is symmetric and peaked at the centre (i.e.\ at the location of the pointing).
For higher energy bins, we observe an increase of the background rate at large offset angles, leading to a ring-shaped distribution at the highest energies.
This is due to the poor rejection power for cosmic-ray background events obtained with the analysis configuration that has been employed to prepare the data used here; we observe this feature to be much less pronounced for an analysis configuration with better gamma-hadron separation \citep[e.g.\ as in][]{Ohm2009}.

\begin{figure}
  \centering
  \includegraphics{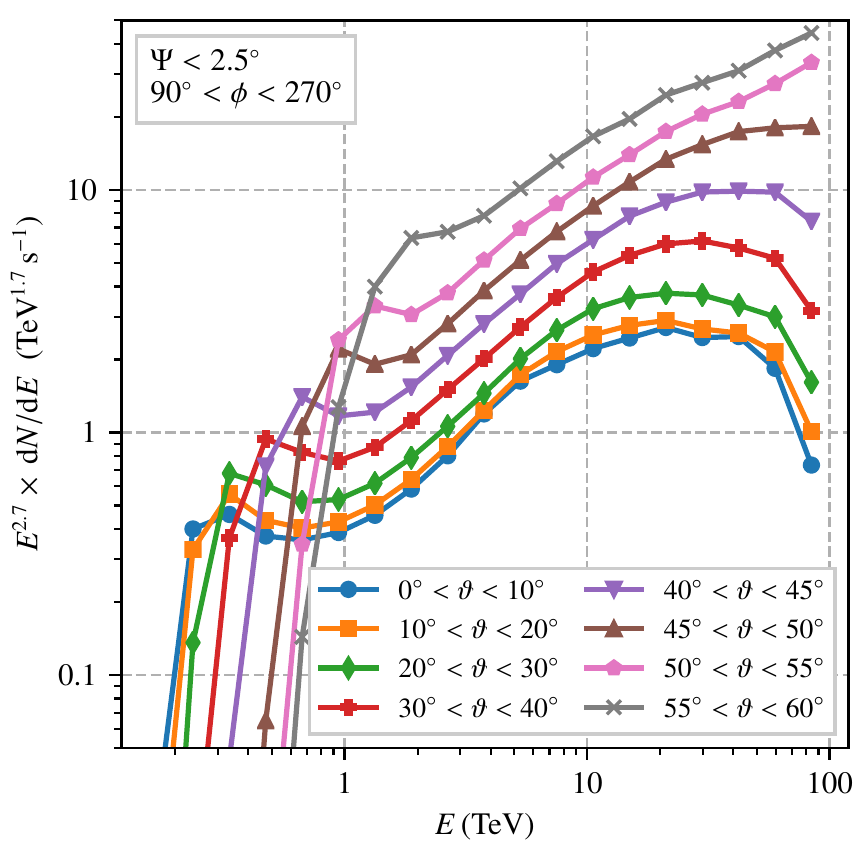}
  \caption{Background model energy spectrum in different bins of zenith angle $\vartheta$, for azimuth angles $90^\circ<\phi<270^\circ$.
  Shown is the rate integrated in a circle around the pointing position with radius $2.5^\circ$.
  To enhance features, the vertical axis is multiplied by $E^{2.7}$.}
  \label{fig:bg_spec_zen}
\end{figure}

\begin{figure}
  \centering
  \includegraphics{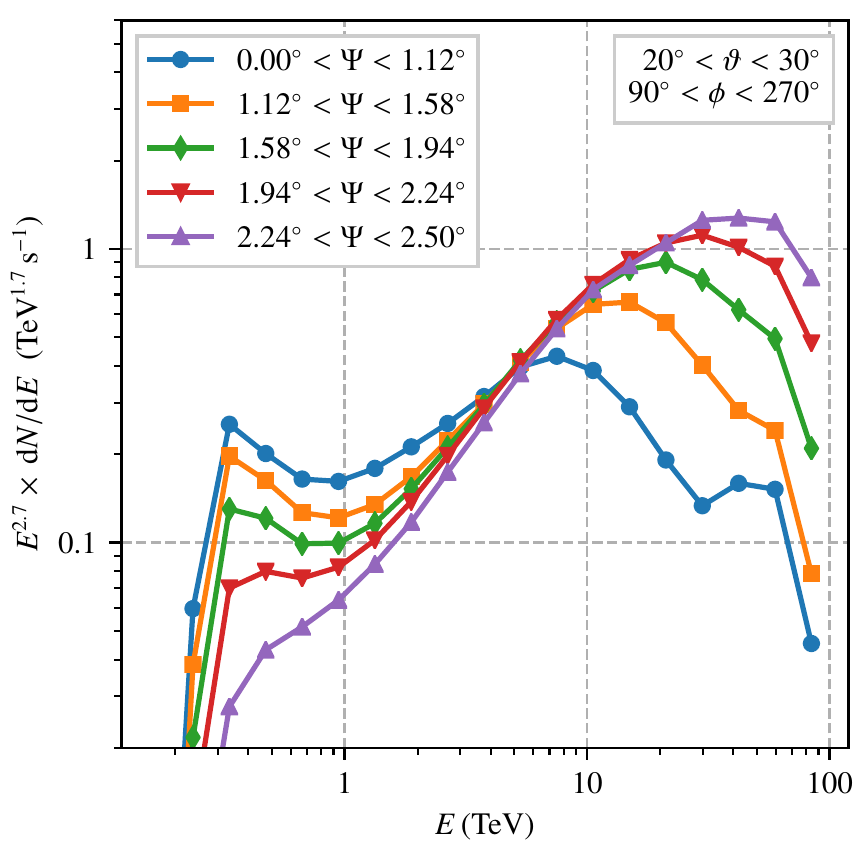}
  \caption{Background model energy spectrum for zenith angles $20^\circ<\vartheta<30^\circ$ and azimuth angles $90^\circ<\phi<270^\circ$.
  Shown is the rate integrated in concentric rings of equal area around the pointing position.
  To enhance features, the vertical axis is multiplied by $E^{2.7}$.}
  \label{fig:bg_spec_off}
\end{figure}

We show the spectral shape of the final model in Figs.~\ref{fig:bg_spec_zen} and~\ref{fig:bg_spec_off} (again before the assignment to a specific observation).
We observe that the spectral shape is close, but not identical, to the shape of the primary cosmic-ray spectrum, which follows a power law $\propto E^{-2.7}$ in good approximation at the energies relevant here.
The discrepancies can be attributed to a dependence of the effective area on the energy, caused for example by an energy-dependent event selection efficiency.

Figure~\ref{fig:bg_spec_zen}, which shows the background model spectrum for different zenith angle bins, illustrates that the energy threshold increases with increasing zenith angle.
This reflects the increased absorption of Cherenkov photons due to, on average, larger distances between the telescopes and the air shower for larger zenith angles.
It is also evident that the effective area of the telescopes increases with increasing zenith angles, leading to a larger rate of background events.
This well-known effect can be understood when considering that air showers that are incident under a large angle illuminate with Cherenkov photons a larger area on the ground, thus increasing the probability that enough telescopes trigger the event.

Figure~\ref{fig:bg_spec_off} shows the background model spectrum for different offset angles~$\Psi$ with respect to the centre of the field of view.
Here, we observe again the feature that, at high energies, the background rate is larger at high offset angles than at the centre.

\subsubsection{Validation}
\begin{figure*}
  \centering
  \includegraphics{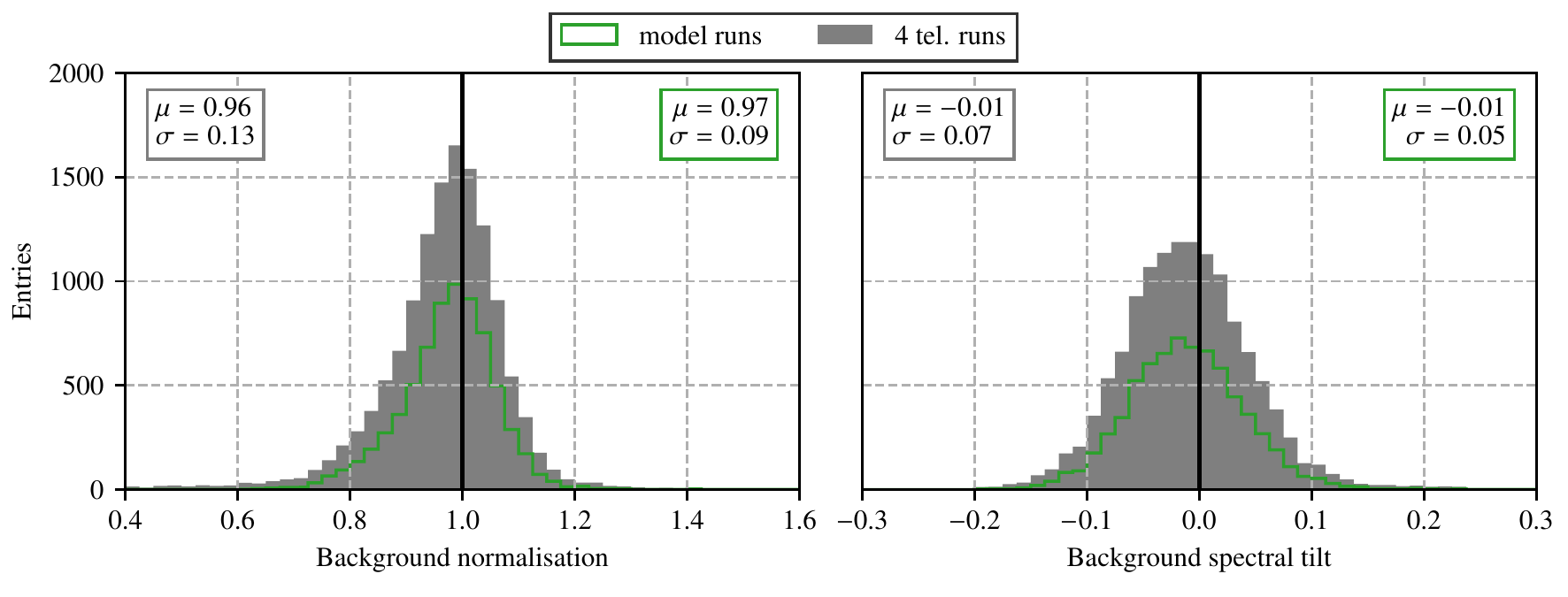}
  \caption{Distribution of fitted background normalisations and spectral tilts ($\delta$, cf.\ Eq.~\ref{eq:tilt}).
  We show results for all 4-telescope observations (grey, filled histograms) and those used in the construction of the background model (green histograms).
  The mean ($\mu$) and standard deviation ($\sigma$) of the distributions are indicated.}
  \label{fig:fit_norm_tilt}
\end{figure*}

Before applying the constructed background model in data ana\-lysis, we performed a general validation of the model by comparing it to archival \hess data.
This procedure is similar to that outlined in Sect.~\ref{sec:refine}, where we already fitted the normalisation of the model to archival observations in separate energy bins.
Here, we adapted this fit such that it resembles more the utilisation of the model in the data analysis with \ct or \gp.
These tools currently offer the possibility to fit a model normalisation (across all energy bins) and a spectral `tilt', that is to say, a parameter $\delta$ that modifies the predicted background rate $R$ at energy $E$ as
\begin{linenomath*}\begin{equation}\label{eq:tilt}
R'=R\cdot (E/E_0)^{-\delta}\,,
\end{equation}\end{linenomath*}
where $E_0=1\,\mathrm{TeV}$ is a reference energy.
Performing a fit of these parameters to all archival \hess Phase-I observations, we obtained the parameter distributions displayed in Fig.~\ref{fig:fit_norm_tilt}.

Considering only observations used in the construction of the model, we obtain again a normalisation distribution with a standard deviation of 9\% (cf.~Fig.~\ref{fig:fit_norms}).
This, together with the narrow distribution obtained for the spectral tilt parameter (width~0.05), demonstrates the validity of the background model in terms of its normalisation and spectral shape for a very large and diverse set of observations.

\begin{figure}
  \centering
  \includegraphics{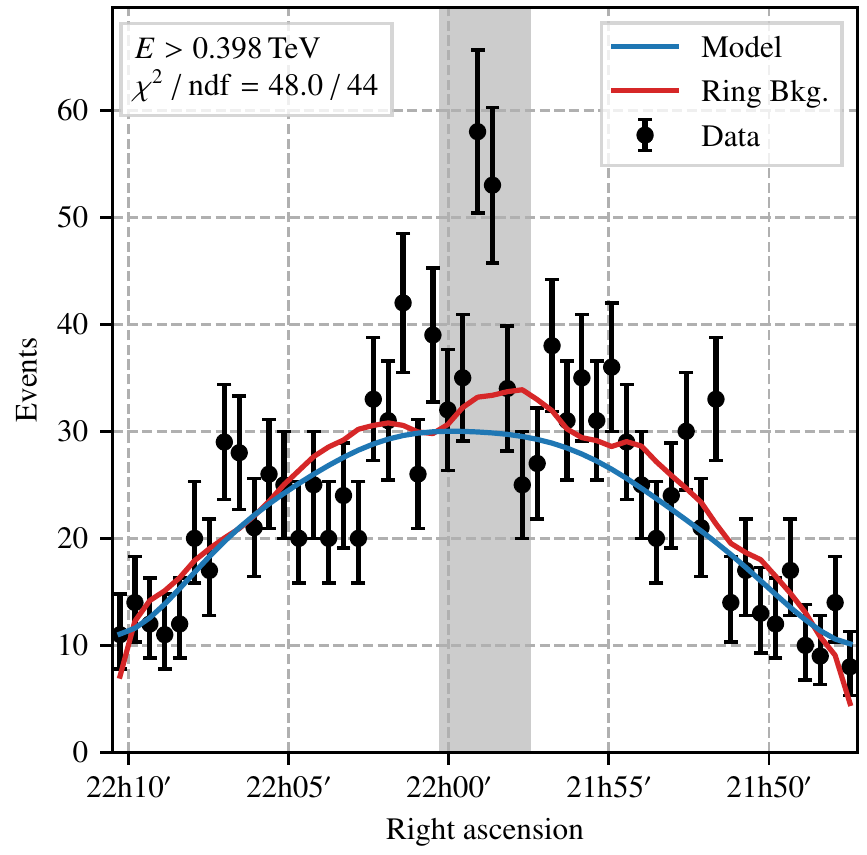}
  \caption{Number of events for observation 47829 at final selection level, as a function of right-ascension (J2000).
  The observation is part of the \pks (steady) data set.
  We show all events with declination $|\delta-\delta_\mathrm{pnt}|<0.7\,\mathrm{deg}$, where $\delta_\mathrm{pnt}$ is the declination coordinate of the pointing position.
  Black points are the measured data; the prediction of the fitted background model is shown as a blue line.
  The grey-shaded area marks an exclusion region around the position of \pks.
  The text box in the upper left corner denotes the applied energy threshold as well as the result of a $\chi^2$ test for the agreement between the model and the measured data.
  We show in addition the number of background events obtained from the ring background method for this observation (red line, cf. Sect.~\ref{sec:ring}).}
  \label{fig:example_proj}
\end{figure}

Evaluating the validity of the prediction of the spatial shape of the model is slightly more complicated, owing to the fact that the spatial shape is not easily parametrisable.
For single observations, the validity can be checked, for instance, by inspecting the predicted and observed rate for one-dimensional slices; an example is shown in Fig.~\ref{fig:example_proj}.
We observe a good agreement between the prediction of the fitted model and the measured data, as well as between the model prediction and the prediction of the ring background method in this case.

\begin{figure*}
  \sidecaption
  \includegraphics{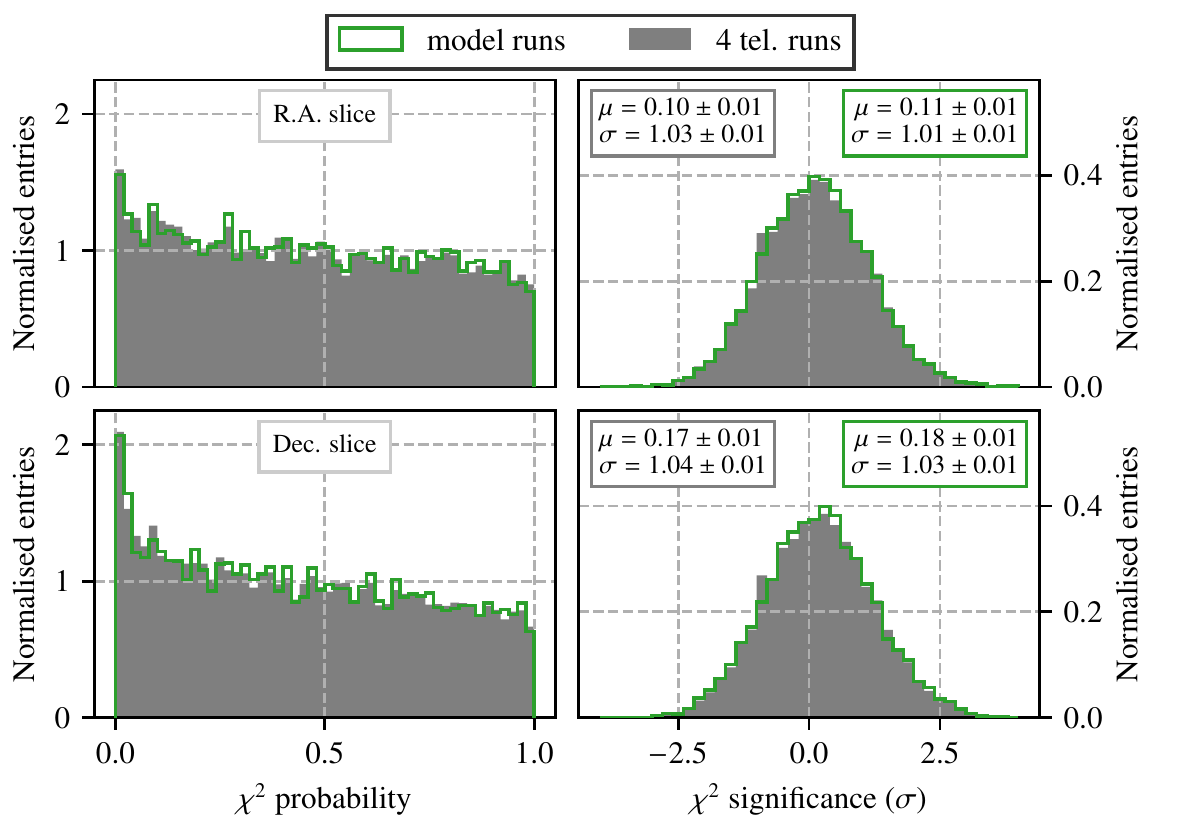}
  \caption{Distributions of $\chi^2$ statistic values.
  We show results for all 4-telescope observations (grey, filled histograms) and those used in the construction of the background model (green histograms).
  Distributions are shown for a slice along the right-ascension (top) and declination (bottom) axis (see main text for details).
  Displayed is the $\chi^2$ probability (left) and the corresponding significance in terms of standard deviations of a normal distribution (right).
  The fitted mean ($\mu$) and width ($\sigma$) of the significance distributions are indicated as well.}
  \label{fig:fit_chi2}
\end{figure*}

For a more quantitative evaluation that can also be applied to many observations, we performed a $\chi^2$ test for these one-dimensional slices, both along the right-ascension and declination axis.
Since the model can only be expected to give a good prediction for regions that are free of \gam-ray sources, we excluded regions that contain known sources from the $\chi^2$ computation.
The $\chi^2$ test yields a p-value that is expected to follow a flat distribution if the model describes the data perfectly.
We show the distributions obtained for slices along both axes in the left panels of Fig.~\ref{fig:fit_chi2}, observing only slight deviations from a flat distribution.
The right panels display distributions of the corresponding significance values in terms of standard deviations of a normal distribution.
Again, we observe only a small deviation from the expectation of a Gaussian distribution with zero mean and unity width, concluding that also the spatial shape of the cosmic-ray background is described well by our model.
Inspecting closer those observations with the smallest $\chi^2$ probabilities, we found that sometimes hardware malfunctions or bad atmospheric conditions are responsible for the bad agreement.
However, we were not able to identify single causes that are responsible for a bad agreement in a majority of the observations.
We note that it is possible to utilise the result of the $\chi^2$ test as an additional quality criterion for the analysis, that is, to discard observations for which the agreement of the background model fit is particularly bad.

\section{Validation of the 3D likelihood analysis}
\label{sec:template}
We applied the 3D likelihood analysis, using \ct and \gp, to the same data sets that we have used for the validation of the analysis tools in Sect.~\ref{sec:standard}.
In each case, we described the cosmic ray-induced background using the background model template introduced in the previous section.
Sect.~\ref{sec:3d_analysis_description} provides a description of the analysis details.
We present the analysis results in Sect.~\ref{sec:3d_analysis_results}.

\subsection{Analysis description}
\label{sec:3d_analysis_description}
The analysis is based on a fit of (typically multiple) models to the observed data.
In this paper, we always use one model (per observation) for the residual cosmic-ray background and one model for the analysed source.
The fit proceeds via an optimisation of a likelihood function that expresses the agreement between the model prediction and the observed data, taking into account the instrument response functions (IRFs), that is, the effective area, the point spread function (PSF), and the energy dispersion matrix.
The likelihood function depends on the reconstructed direction and energy of the observed events, meaning that it has three dimensions.

\subsubsection{General settings}
We performed the likelihood analysis with the two tools in a conceptionally different way: while we carried out an unbinned likelihood fit with \ct, we used a binned likelihood fit in \gp.
The former uses probability density functions for the IRFs to determine a likelihood for each observed event, while the latter is a forward-folding fit employing a Poisson likelihood in each bin.
The reason for this choice is that, on the one hand, an unbinned fit is not possible yet with \gp, while on the other hand, a binned fit including full treatment of the energy dispersion is computationally extremely intensive with the current version of \ct at least for parts of the data sets analysed here.
Both methods yield identical results for large statistics and not-too-coarse binning.

With both tools, we employed a `joint' fit, that is, we calculated a likelihood value for each observation of the data set and multiplied these values to obtain the final likelihood value.
This is opposed to a `stacked' fit, for which the measured data are summed over all observations and a model prediction for the entire data set, based on averaged IRFs, is obtained, leading to only one likelihood value.
The joint fit, not relying on an averaged description of the instrument, is generally expected to lead to a more accurate model for the observed data.
However, since there is typically at least one free fit parameter for the background model template of each observation in a joint fit, it can have many more free parameters than the corresponding stacked fit for the same data set.
This is still feasible for the data sets analysed here, where the maximum number of observations in a data set is~20 (for \msh).
We expect that for considerably larger data sets, the fit becomes computationally prohibitive due to the large number of free parameters, calling for more elaborate analysis strategies.

For the binned fit with \gp, as well as for the generation of result sky maps, we utilised spatial pixels with $0.02^\circ$ side length.
The energy axis is defined with eight logarithmically spaced bins per decade in energy.

In the analysis of IACT data, the largest systematic uncertainties in the description of the residual cosmic-ray background typically occur at large offset angles, at the edges of the observed field of view, and at the lowest energies, close to the instrument threshold.
We therefore restricted the analysis to events within a maximum offset angle, \psimax, and above an energy threshold value, \ethr.
We chose \psimax such that the observed source as well as sufficiently large areas free of known \gam-ray sources (the latter providing constraints for the background model template) are enclosed inside the selected region for all observations.
We computed the energy threshold for each observation separately, defining it as $E_\mathrm{thr}=\max(E_\mathrm{thr}^\mathrm{bias}, E_\mathrm{thr}^\mathrm{bkg})$.
Here, $E_\mathrm{thr}^\mathrm{bias}$ denotes a threshold value that ensures that the bias of the energy reconstruction does not exceed 10\% for events within the maximum offset angle; this requirement is similar to that imposed in the spectrum extraction with the reflected background method (cf.~Sect.~\ref{sec:reflected}).
In addition, we introduced a threshold value $E_\mathrm{thr}^\mathrm{bkg}$ for the background model of each observation, conservatively defined as the upper edge of the energy bin with the highest predicted background rate.
This was necessary here because many of the analysed observations have been taken very early in the operation time of \hess, when the optical efficiency of the telescopes was very high, leading to a comparatively low $E_\mathrm{thr}^\mathrm{bias}$.
These observations are sometimes not described well close to the threshold by the background model, which is constructed from observations that, on average, have been conducted with lower optical efficiency of the telescopes.
Thus, $E_\mathrm{thr}^\mathrm{bkg}$ is larger than $E_\mathrm{thr}^\mathrm{bias}$ for most of the observations analysed here.
This is not the case in analyses of data sets recorded under less exceptional conditions, which can likely proceed without this additional restriction.
We list the resulting range of values for \ethr, as well as the employed values of \psimax, for each data set in Table~\ref{tab:template_cuts}.

\begin{table}
  \caption{Event selection criteria for 3D likelihood analysis.}
  \label{tab:template_cuts}
  \centering
  \begin{tabular}{ccc}
    \hline
    Source & \psimax & \ethr\\
     & (deg) & (TeV)\\\hline
    \crab & 2.0\tablefootmark{a} & 0.750 -- 0.825\\
    \pks & 1.5 & 0.422 -- 0.619\\
    \msh & 1.5 & 0.422 -- 0.562\\
    \rxj & 1.5\tablefootmark{b} & 0.316 -- 0.422\\
    \hline
  \end{tabular}
  \tablefoot{\psimax denotes the maximum event offset angle;
  the last column gives the minimum and maximum applied energy threshold of all observations in the data set.\\
  \tablefoottext{a}{We use the larger value of $2^\circ$ for the \crab because two (out of four) observations of this data set have been taken with an offset angle of $1.5^\circ$ w.r.t.\ the source position.}
  \tablefoottext{b}{The data set for \rxj contains one observation with a considerably larger offset between the pointing direction and the source w.r.t.\ the remaining observations.
  For this observation, the source is not fully enclosed in the selected region.
  This leads to a slight loss in sensitivity, which we accept here for the sake of a consistent handling of the observations in the analysis.}
  }
\end{table}

\subsubsection{Background and source models}
\label{sec:bg_src_templates}
As model for the residual cosmic-ray induced background, we used the background model template developed in Sect.~\ref{sec:background}.
This template, specific for each observation, can be modified in the fit by two parameters: a global normalisation factor, and a spectral `tilt', as defined previously in Eq.~\ref{eq:tilt}.

Like in Sect.~\ref{sec:reflected}, we always used a simple power law as spectral model for the source, with the flux normalisation and spectral index as free fit parameters (see Eq.~\ref{eq:powerlaw}).
For the estimation of flux points, we re-performed the analysis for each of the energy bins, fixing the spectral index, the background tilt parameters, and the parameters of the spatial source model to their best-fit values, but leaving free the background and source normalisation parameters.

The different morphologies of the sources analysed in this paper call for different spatial source models.
For the \crab and \pks, we employed the model of a point-like source, with the two source coordinates as free parameters.
The morphology of \msh is still simple enough to be able to use an analytical spatial model for this source as well; we used an elliptical disk model with five free parameters (the source coordinates and the major axis, eccentricity, and position angle of the ellipse) here.
In contrast, the complex morphology of \rxj (cf.~Fig.~\ref{fig:map_rxj1713}) prohibits the use of an analytical model.
We have therefore developed a procedure to generate an `excess template' as spatial source model.
As the name suggests, this template represents a map of the excess of events that can be attributed to \gam-ray emission from the source.
We derived it by fitting only the background model template to the observations, excluding a region around the source from the fit, and subtracting the resulting best-fit background model from the observed data.
We note that the template -- being derived from the data themselves -- is subject to the same statistical fluctuations as the observed data, implying that this approach can in principle lead to a bias of the fitted parameters.
In an attempt to minimise such a bias, we smoothed the excess map using a two-dimensional cubic spline function, thus reducing the statistical fluctuations.
Finally, we clipped the derived template map at zero, removing negative entries.
We furthermore note that, since the excess template is derived from observed data, it need not be convolved with the PSF in the fit; this is possible currently with \gp but not with \ct.
We performed an analysis with the excess template approach not only for \rxj, but also for \msh, being able to compare to the results obtained with the elliptical disk model in that case.

\subsection{Results}
\label{sec:3d_analysis_results}
\begin{figure}
  \centering
  \subfigure[Flux normalisation $\phi$.]{
    \includegraphics{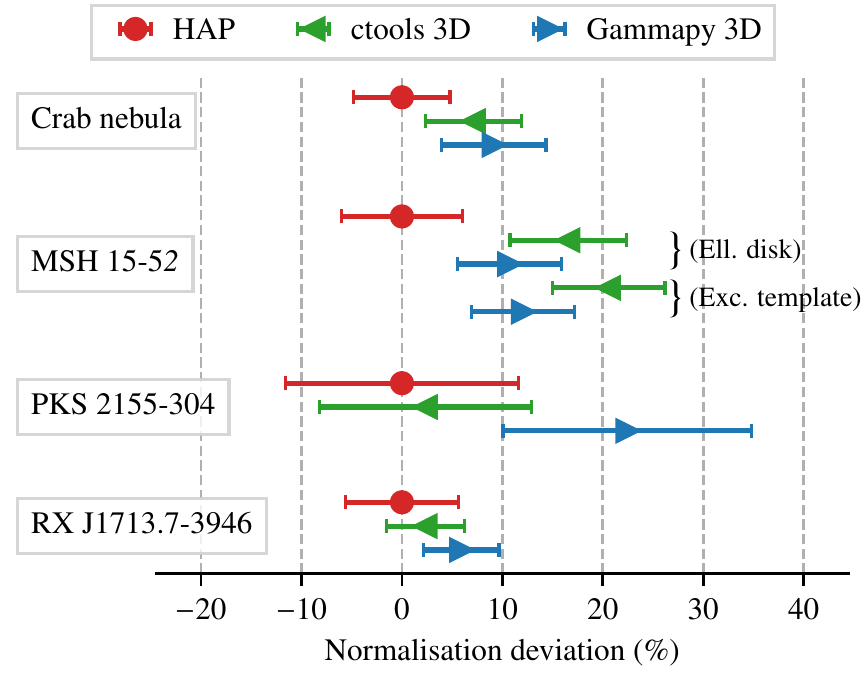}
    \label{fig:pl_norm_dev_3d}
  }\\
  \subfigure[Spectral index $\Gamma$.]{
    \includegraphics{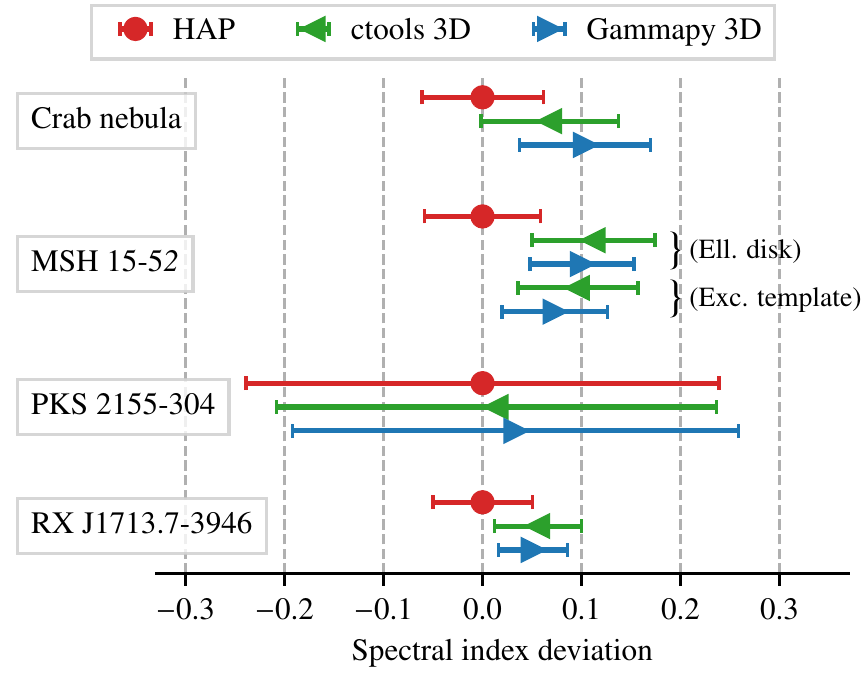}
    \label{fig:pl_index_dev_3d}
  }
  \caption{Comparison of spectral fit parameters for all sources, obtained using 3D likelihood analysis with \ct and \gp.
  We plot the deviation w.r.t.\ the results obtained with the reflected background method and our standard tool \hap.
  The error bars denote statistical uncertainties only (68\% confidence level).}
  \label{fig:pl_par_dev_3d}
\end{figure}

We summarise the results of all 3D likelihood analyses in Table~\ref{tab:fitresults}.
Furthermore, Fig.~\ref{fig:pl_par_dev_3d} shows a comparison of the best-fit parameter values of the power law model for all sources.
The spectrum obtained with the 3D likelihood analysis is steeper (i.e.\ the spectral index $\Gamma$ is larger) than that obtained with the reflected background method in all cases.
Comparing the spectra in detail, we find that this is partly due to an improved sensitivity of the 3D likelihood analysis at high energies, where we obtain only flux upper limits because the excess of \gam-ray events is not significant.
Here, the reflected background method suffers from poor statistics in the off regions as well, leading to an inaccurate background estimate and hence bad sensitivity.
In contrast, the cosmic-ray background model employed in the 3D likelihood analysis, being derived from many observations, is afflicted less by this problem.
Consequently, the 3D likelihood analysis yields slightly steeper spectra, and more constraining upper limits at high energies.
Furthermore, the fact that we apply slightly higher energy thresholds in the 3D likelihood analyses can also explain part of the discrepancy, in particular if the intrinsic source spectrum is not a true power law.
We note that the deviation between the two methods is at most $\sim 0.1$ for the spectral index and $\sim 20\%$ for the flux normalisation; this is within the systematic uncertainties on these parameters that are usually quoted by \hess \citep[see e.g.][]{hesscrab2006}.

In the following, the results obtained for the different data sets are discussed in detail.
Due to space restrictions, we can show spectra and maps only for a selection of the analyses here; all remaining plots can be found in Appendix~\ref{sec:appendix_template}.
Furthermore, we always only show maps derived with one of the open-source tools, implying that the corresponding maps derived with the other tool are qualitatively and quantitatively compatible.

\subsubsection{\crab and \pks}
\begin{figure}
  \centering
  \includegraphics{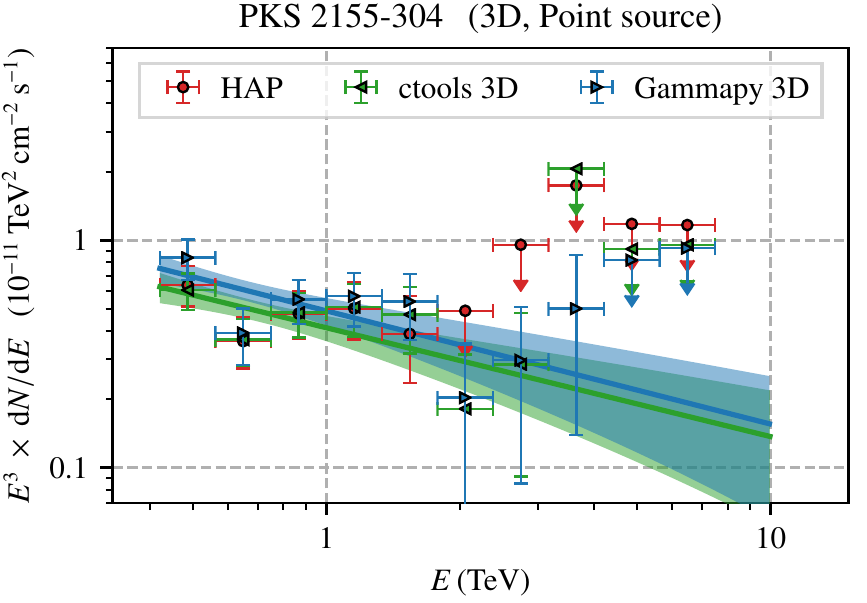}
  \caption{Comparison of spectra for \pks.
  The spectra shown in green and blue were derived using a 3D likelihood analysis with \ct and \gp, respectively.
  The butterflies show the fitted power laws.
  The results are compared to those obtained with the reflected background method using the \hap software (in red).
  We compute upper limits (95\% confidence level) for flux points with a statistical significance of less than two standard deviations.}
  \label{fig:spectrum_3d_pks2155}
\end{figure}

\begin{figure}
  \centering
  \includegraphics{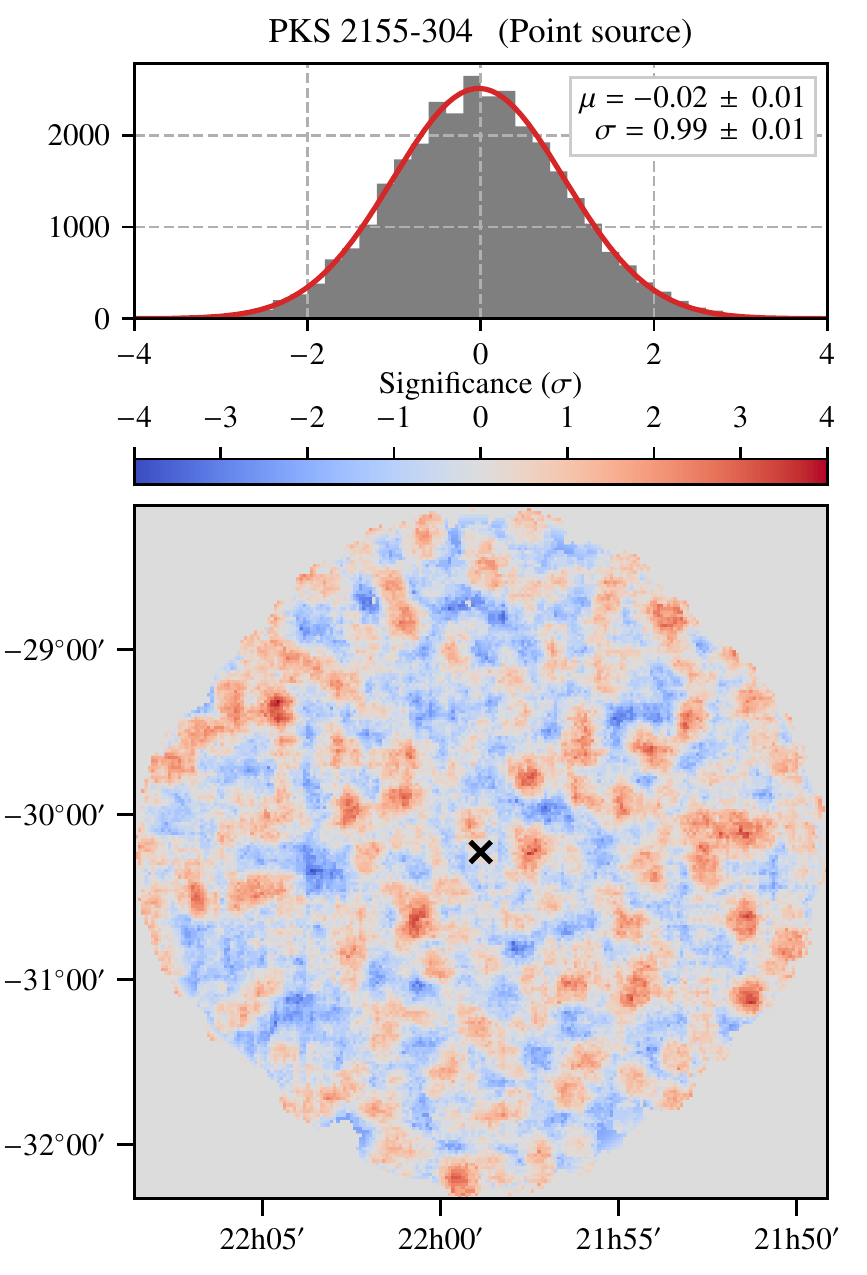}
  \caption{Residual significance map for \pks, in equatorial coordinates (J2000).
  Results derived with the 3D likelihood analysis with \gp are displayed.
  We apply a convolution with a top-hat kernel of $0.1^\circ$ radius to reduce statistical fluctuations.
  The position of \pks is indicated by the `$\times$'.
  The upper panel shows the distribution of significance values, together with the fit of a normal distribution (red line).}
  \label{fig:sign_map_3d_pks2155}
\end{figure}

We show the energy spectra that we obtain for \pks and the \crab in Figs.~\ref{fig:spectrum_3d_pks2155} and~\ref{fig:spectrum_3d_crab}, respectively.
Figs.~\ref{fig:sign_map_3d_pks2155} and~\ref{fig:sign_map_3d_crab} show the respective residual significance maps, together with the corresponding entry distributions.
The computation of the significance maps starts by convolving the best-fit model prediction and the observed data (both integrated over energy) with a top-hat kernel of $0.1^\circ$ radius, thus smoothing statistical fluctuations (cf.~Sect.~\ref{sec:ring}).
We then calculate the significance of the observed data under the hypothesis of the best-fit model following \citet{Li1983}, adopting the limiting case of a perfectly known number of `off' counts (corresponding to the best-fit model prediction in our case).

The data sets of the \crab and \pks comprise few observations and are hence dominated by statistical rather than systematic uncertainties.
Hence, unsurprisingly, the likelihood analysis works well for these data sets, yielding results that are highly compatible with those obtained with standard analysis techniques.
The significance maps are governed entirely by statistical fluctuations, indicating an almost perfect description of the analysed source as well as the residual background in the field of view.

\subsubsection{\msh}
\begin{figure*}
  \centering
  \includegraphics{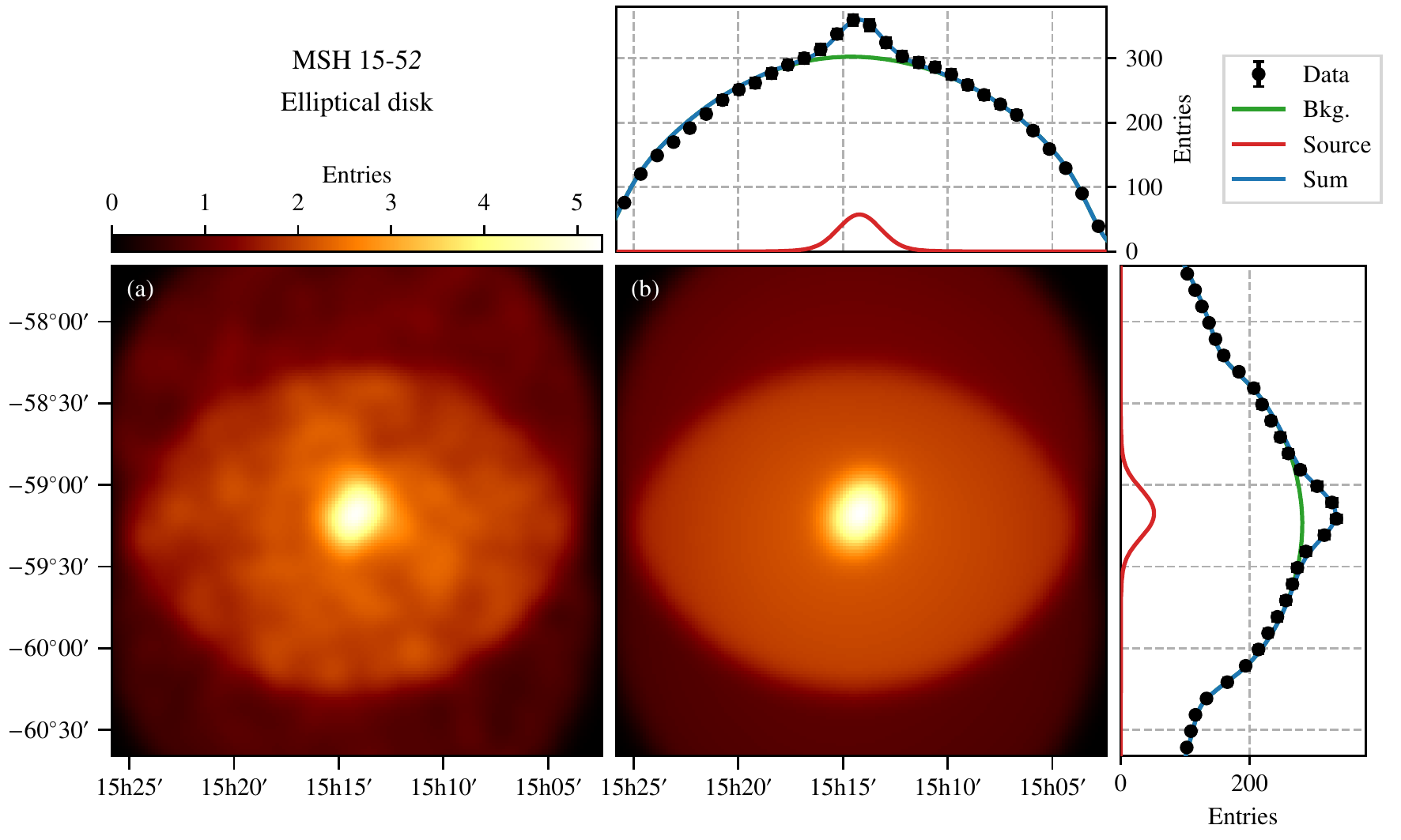}
  \caption{Counts map (a) and best-fit model map (b) for \msh, in equatorial coordinates (J2000).
  Results derived with the 3D likelihood analysis with \ct using an elliptical disk as source model are displayed.
  The maps have been integrated over all energy bins contributing to the fit and smoothed using a Gaussian kernel with a width of $0.08^\circ$.
  The small panels show projections onto the two spatial axes; the separate components are indicated as well here.}
  \label{fig:data_model_map_3d_msh1552}
\end{figure*}

Figure~\ref{fig:data_model_map_3d_msh1552} shows sky maps of the observed data as well as the best-fit model obtained with the 3D likelihood analysis for \msh, using the elliptical disk as spatial model for the source.
We smoothed the maps using a Gaussian kernel with a width of $0.08^\circ$, which approximately corresponds to the size of the PSF for the data sets analysed here.
We observe a very good agreement between the two maps, indicating that both the cosmic-ray background and \msh are described well by the fitted models.
The slight disagreement between the background model prediction and the data in the western part of the maps does not seem to affect the fitted source model.
This interpretation is supported by the essentially featureless significance map for this analysis, which we show in Fig.~\ref{fig:sign_map_3d_msh1552} in Appendix~\ref{sec:appendix_template}.

\begin{figure}
  \centering
  \includegraphics{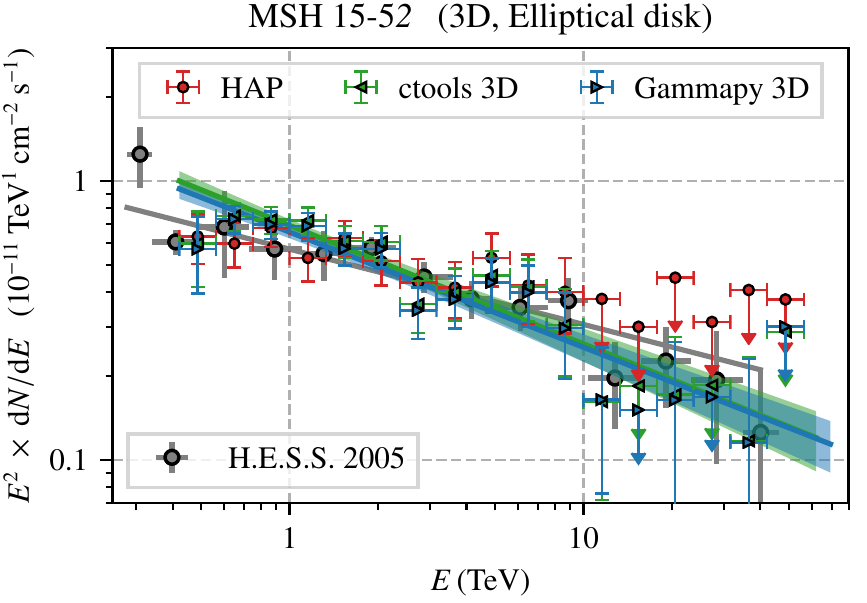}
  \caption{Comparison of spectra for \msh.
  The spectra shown in green and blue were derived using a 3D likelihood analysis with \ct and \gp, respectively, employing an elliptical disk as source model.
  The butterflies show the fitted power laws.
  The results are compared to those obtained with the reflected background method using the \hap software (in red).
  We compute upper limits (95\% confidence level) for flux points with a statistical significance of less than two standard deviations.
  The published spectrum is taken from \citet{hessmsh1552_2005}.}
  \label{fig:spectrum_3d_msh1552}
\end{figure}

We show the spectrum obtained for \msh with the 3D likelihood analysis using an elliptical disk model in Fig.~\ref{fig:spectrum_3d_msh1552}.
In accordance with the maps, we observe an excellent agreement with the spectrum derived with the reflected background method, as well as with the published spectrum \citep{hessmsh1552_2005}.

We show the counts and model maps, the significance map, and the spectrum derived for \msh using the 3D likelihood analysis with an excess template rather than an elliptical disk model in Figs.~\ref{fig:data_model_map_3d_msh1552_excess}, \ref{fig:sign_map_3d_msh1552_excess}, and \ref{fig:spectrum_3d_msh1552_excess}, respectively.
The results are highly compatible with those obtained with the elliptical disk model, giving us confidence that the procedure of generating a model template from the excess map is valid and can also be applied to the analysis of \rxj. 

\subsubsection{\rxj}
\begin{figure}
  \centering
  \includegraphics{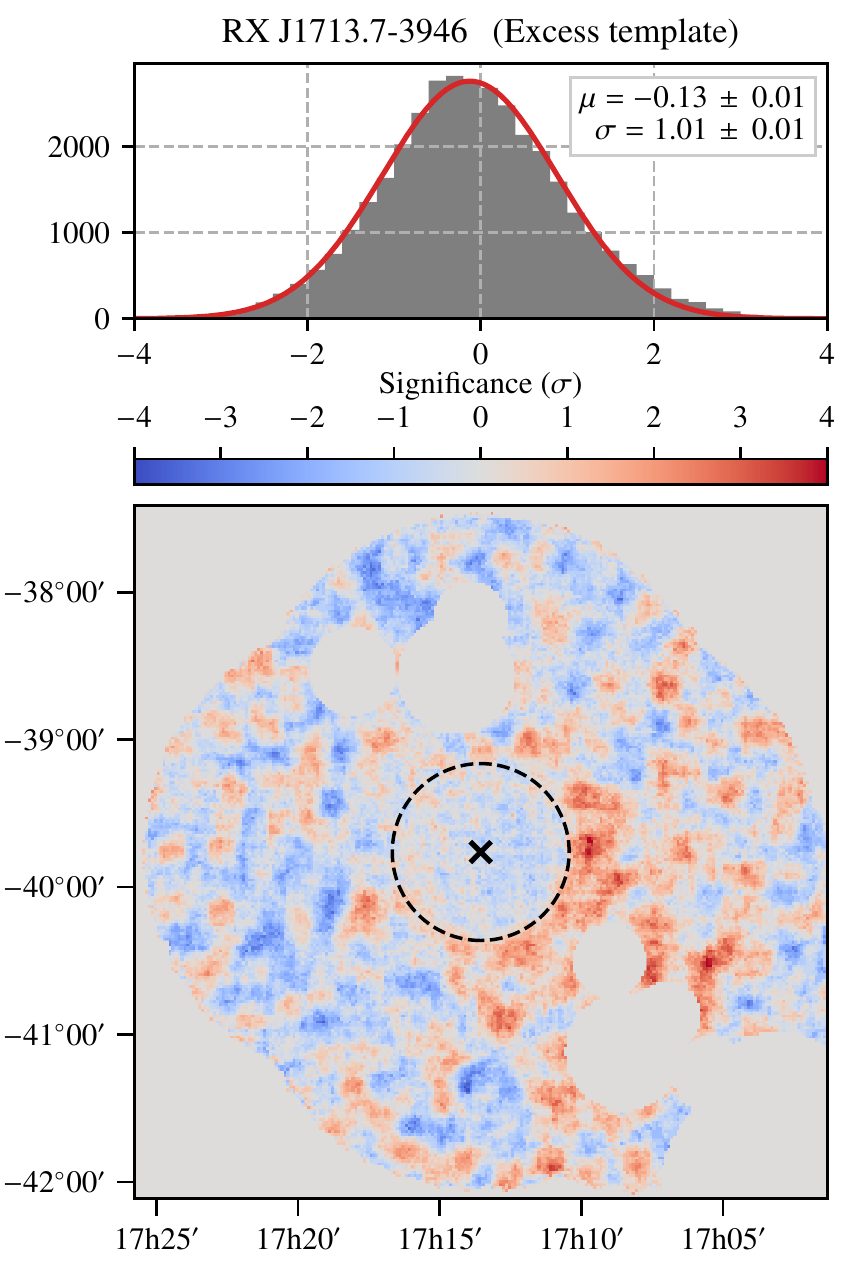}
  \caption{Residual significance map for \rxj, in equatorial coordinates (J2000).
  Results derived with the 3D likelihood analysis with \gp are displayed.
  The position of \rxj is indicated by the `$\times$'; the dashed line shows the size of the excess template.
  The upper panel shows the distribution of significance values, together with the fit of a normal distribution (red line).}
  \label{fig:sign_map_3d_rxj1713}
\end{figure}

Figure~\ref{fig:sign_map_3d_rxj1713} shows the significance map derived for the 3D likelihood analysis of \rxj.
The grey regions contain known \gam-ray sources or bright stars and were masked in the fit\footnote{
In principle, the likelihood analysis offers the possibility to model all sources in the field of view.
A full analysis of the field of view containing \rxj is however beyond the scope of this paper.}\footnote{
The exclusion of sub-regions from the fit is not easily possible in an unbinned fit with the version of \ct that we have used.
Hence, no regions were excluded in the \ct fit.
Performing the \gp fit without exclusion regions as well, we find that the results are altered by less than 5\%.}.
We observe two noticeable features.
First, there is a residual positive excess to the south-east of \rxj.
This could be caused by a true excess of \gam rays (e.g.\ from unresolved sources, or of diffuse nature) as well as by an imperfect model of the cosmic-ray background, likely a combination of both.
Second, the region covered by the source model template is almost free of statistical fluctuations.
This is an artefact of the generation of the source model from the excess map (cf.~Sect.~\ref{sec:bg_src_templates}); although we applied a smoothing algorithm, the excess template necessarily is subject to the same statistical fluctuations as the data it is derived from.

Finally, we refer to the data and model maps and the spectrum for \rxj in Figs.~\ref{fig:data_model_map_3d_rxj1713} and~\ref{fig:spectrum_3d_rxj1713}, respectively.
Similarly as for the results obtained with the reflected background method, we observe a disagreement to the published spectrum \citep{hessrxj1713_2018} at energies below $\sim 0.45\,\mathrm{TeV}$, that is, for the first derived flux point.
The disagreement is reduced with respect to the standard method, indicating that the 3D likelihood analysis is a better choice of analysis method for this source.
That a deviation to the published spectrum remains is likely due to a systematic problem with the (relatively small) data set analysed here.


\section{Conclusion}
\label{sec:conclusion}
In this paper, we demonstrate the applicability of the open-source high-level \gam-ray analysis tools \ct and \gp to experimental data recorded by the \hess system of Cherenkov telescopes.
In Sect.~\ref{sec:standard}, we show that when applying standard, well-tested analysis techniques, \ct and \gp are able to exactly reproduce results obtained with the proprietary \hess analysis software.
We then focus on the 3D likelihood analysis, an analysis approach that is new in the field of very-high-energy \gam-ray astronomy.
First, in Sect.~\ref{sec:background}, we introduce a procedure to construct a model template for the residual cosmic ray-induced background, one of the key ingredients for the 3D likelihood analysis approach.
Furthermore, we characterise the principal features of the resulting background model and perform a general validation.
Finally, in Sect.~\ref{sec:template}, we apply the 3D likelihood analysis to \hess data.
We obtain results that are highly compatible with those derived with standard analysis techniques, thus demonstrating the validity of the background model as well as the analysis approach itself.

We note that the data set used for analysis verification in this paper (cf.~Sect.~\ref{sec:hess}) has various limitations:
it comprises only few observations taken on relatively strong \gam-ray sources; the corresponding fields of view do not require the modelling of multiple source components; and the analysis configuration utilised to process the data is no longer state-of-the-art.
However, we are confident that the analysis concept can also be applied to larger data sets, more intricate fields of view, and data processed with up-to-date analysis configurations; first successful attempts have been made by \citet{mayer2014}, \citet{devin2018}, and \citet{ziegler2018}.
Furthermore, we remark that particularly complicated sky regions, such as for example the Galactic centre region, cannot be properly analysed with traditional analysis techniques at all, calling for new approaches like the one presented here.
Nevertheless, we expect that further studies will be necessary for this.
For instance, the question up to which level of precision the residual cosmic-ray background can be modelled using the approach described here for deep observations is important, but beyond the scope of this paper.

That we have successfully validated the application of \ct and \gp to the analysis of \hess data is important not only for the \hess experiment, but also for the upcoming Cherenkov Telescope Array.
First, it shows that the development of the analysis tools that could be used for CTA is progressing well; both packages -- while still in development -- can already be considered mature now.
Second, this work also paves the way for the application of the 3D likelihood analysis to CTA data.
CTA will have greatly improved sensitivity with respect to current instruments and is thus expected to discover many new sources of \gam rays.
Therefore, it will benefit from the 3D likelihood analysis approach, which is designed to simultaneously analyse multiple components in the observed field of view.

\newpage
\begin{acknowledgements}
We would like to thank the \hess Collaboration for allowing us to use their internal analysis software package and their archival data; we used the latter to construct and validate the background model.
Furthermore, we acknowledge the preparatory work of the members of the \hess FITS task group, which have prepared the \hess public test data release.
We thank the developer teams of the \ct and \gp packages for helpful support during the preparation of this paper.
This research made use of ctools, a community-developed analysis package for Imaging Air Cherenkov Telescope data.
ctools is based on GammaLib, a community-developed toolbox for the scientific analysis of astronomical gamma-ray data.
This research made use of Gammapy, a community-developed core Python package for gamma-ray astronomy.
This research made use of Astropy\footnote{\url{http://www.astropy.org}}, a community-developed core Python package for Astronomy \citep{astropy:2013,astropy:2018}.
The plots shown in this paper have been produced with the matplotlib\footnote{\url{http://www.matplotlib.org}} package \citep{Hunter:2007}.
The authors acknowledge support by the German Federal Ministry of Education and Research under grant ID 05A17WE1.
The authors gratefully acknowledge the compute resources and support provided by the Erlangen Regional Computing Center (RRZE).
\end{acknowledgements}

\bibliographystyle{aa}
\vspace{3cm}
\bibliography{references}


\begin{appendix}
\section{Reflected background method spectra for point sources}
\label{sec:appendix_reflected}
\begin{figure}
  \centering
  \includegraphics{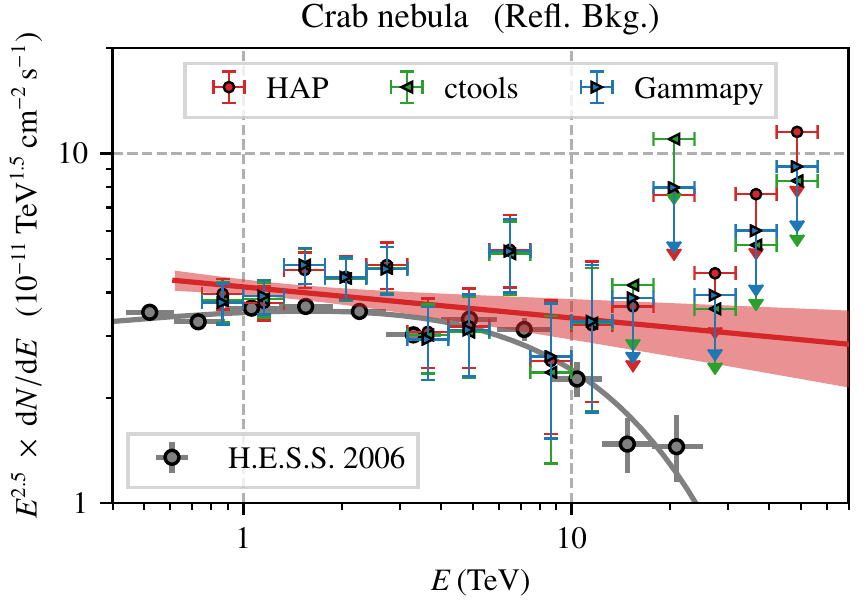}
  \caption{Comparison of spectra for the \crab.
  The spectra shown in red, green, and blue were derived with the reflected background method with \hap, \ct, and \gp, respectively.
  For the \hap analysis, we show the result of the power-law fit in addition.
  We compute upper limits (95\% confidence level) for flux points with a statistical significance of less than two standard deviations.
  The published spectrum is taken from \citet{hesscrab2006}; it uses a power law with exponential cut-off as spectral model.}
  \label{fig:spectrum_standard_crab}
\end{figure}

\begin{figure}
  \centering
  \includegraphics{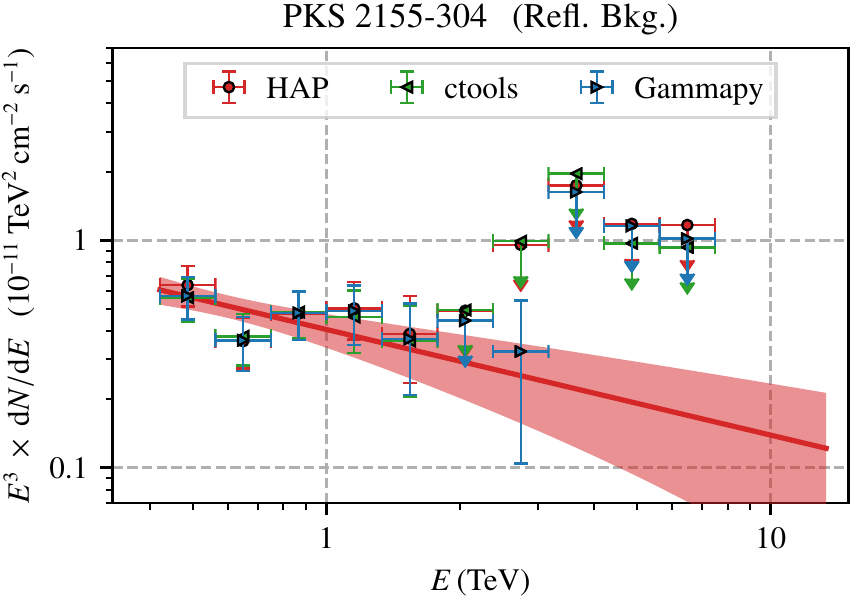}
  \caption{Comparison of spectra for \pks.
  The spectra shown in red, green, and blue were derived with the reflected background method with \hap, \ct, and \gp, respectively.
  For the \hap analysis, we show the result of the power-law fit in addition.
  We compute upper limits (95\% confidence level) for flux points with a statistical significance of less than two standard deviations.}
  \label{fig:spectrum_standard_pks2155}
\end{figure}

We show the spectra extracted with the reflected background method for the \crab and \pks in Figs.~\ref{fig:spectrum_standard_crab} and \ref{fig:spectrum_standard_pks2155}, respectively.
Since \pks is a variable source, we cannot meaningfully compare our results with the literature in this case.

\section{Fit results}
\label{sec:appendix_fit_results}
Table~\ref{tab:fitresults} lists the results of the spectral fits performed with the classical analysis approach (reflected background method, cf.~Sect.~\ref{sec:standard}), as well as the results obtained with the 3D likelihood analyses described in Sect.~\ref{sec:template} (marked with the annotation ``3D'' in the table).

\begin{sidewaystable*}
  \caption{Comparison of best-fit parameter values obtained in all analyses.}
  \label{tab:fitresults}
  \centering
  \begin{tabular}{cccccccccc}
    \hline
    Analysis & Source model & R.A. (J2000) & Dec. (J2000) & $a$ & $e$ & P.A. & $E_0$ & $\phi$ & $\Gamma$ \\
     & & (deg) & (deg) & (deg) & & (deg) & (TeV) & ($10^{-12}\,\mathrm{cm}^{-2}\,\mathrm{s}^{-1}\,\mathrm{TeV}^{-1}$) & \\\hline
    \multicolumn{10}{c}{}\\
    \multicolumn{10}{c}{}\\
    \multicolumn{10}{c}{\textbf{\crab}}\\\hline
    \hap & -- & -- & -- & -- & -- & -- & 1.45 & $15.9\pm 0.8$ & $2.59\pm 0.06$ \\
    \ct & -- & -- & -- & -- & -- & -- & 1.45 & $16.2\pm 0.9$ & $2.64\pm 0.08$ \\
    \gp & -- & -- & -- & -- & -- & -- & 1.45 & $16.3\pm 0.9$ & $2.63\pm 0.07$ \\
    \ct 3D & Point source & $83.619\pm 0.003$ & $22.025\pm 0.002$ & -- & -- & -- & 1.45 & $17.0\pm 0.8$ & $2.66\pm 0.07$ \\
    \gp 3D & Point source & $83.619\pm 0.003$ & $22.025\pm 0.003$ & -- & -- & -- & 1.45 & $17.3\pm 0.8$ & $2.69\pm 0.07$ \\\hline
    \multicolumn{10}{c}{}\\
    \multicolumn{10}{c}{}\\
    \multicolumn{10}{c}{\textbf{\pks}}\\\hline
    \hap & -- & -- & -- & -- & -- & -- & 0.65 & $18.0\pm 2.1$ & $3.46\pm 0.24$ \\
    \ct & -- & -- & -- & -- & -- & -- & 0.65 & $17.4\pm 2.1$ & $3.52\pm 0.28$ \\
    \gp & -- & -- & -- & -- & -- & -- & 0.65 & $17.5\pm 2.1$ & $3.46\pm 0.25$ \\
    \ct 3D & Point source & $329.725\pm 0.007$ & $-30.227\pm 0.006$ & -- & -- & -- & 0.65 & $18.5\pm 1.9$ & $3.48\pm 0.22$ \\
    \gp 3D & Point source & $329.728\pm 0.006$ & $-30.225\pm 0.006$ & -- & -- & -- & 0.65 & $22.1\pm 2.2$ & $3.50\pm 0.23$ \\\hline
    \multicolumn{10}{c}{}\\
    \multicolumn{10}{c}{}\\
    \multicolumn{10}{c}{\textbf{\msh}}\\\hline
    \hap & -- & -- & -- & -- & -- & -- & 1.4 & $2.63\pm 0.16$ & $2.31\pm 0.06$ \\
    \ct & -- & -- & -- & -- & -- & -- & 1.4 & $2.84\pm 0.16$ & $2.34\pm 0.07$ \\
    \gp & -- & -- & -- & -- & -- & -- & 1.4 & $2.70\pm 0.16$ & $2.35\pm 0.06$ \\
    \ct 3D & Elliptical disk & $228.553\pm 0.011$ & $-59.174\pm 0.006$ & $0.194\pm 0.010$ & $0.794\pm 0.044$ & $151\pm 5$ & 1.4 & $3.07\pm 0.15$ & $2.42\pm 0.06$ \\
    \gp 3D & Elliptical disk & $228.549\pm 0.011$ & $-59.171\pm 0.007$ & $0.196\pm 0.007$ & $0.801\pm 0.038$ & $151\pm 6$ & 1.4 & $2.91\pm 0.14$ & $2.41\pm 0.05$ \\
    \ct 3D & Excess template & -- & -- & -- & -- & -- & 1.4 & $3.17\pm 0.15$ & $2.41\pm 0.06$ \\
    \gp 3D & Excess template & -- & -- & -- & -- & -- & 1.4 & $2.95\pm 0.13$ & $2.39\pm 0.05$ \\\hline
    \multicolumn{10}{c}{}\\
    \multicolumn{10}{c}{}\\
    \multicolumn{10}{c}{\textbf{\rxj}}\\\hline
    \hap & -- & -- & -- & -- & -- & -- & 1.15 & $12.6\pm 0.7$ & $2.16\pm 0.05$ \\
    \ct & -- & -- & -- & -- & -- & -- & 1.15 & $13.5\pm 0.7$ & $2.18\pm 0.05$ \\
    \gp & -- & -- & -- & -- & -- & -- & 1.15 & $12.7\pm 0.7$ & $2.16\pm 0.04$ \\
    \ct 3D & Excess template & -- & -- & -- & -- & -- & 1.15 & $12.9\pm 0.5$ & $2.22\pm 0.04$ \\
    \gp 3D & Excess template & -- & -- & -- & -- & -- & 1.15 & $13.3\pm 0.5$ & $2.21\pm 0.04$ \\\hline
  \end{tabular}
  \tablefoot{
    $E_0$, $\phi$, and $\Gamma$ are parameters of the power law model as defined in Eq.~\ref{eq:powerlaw}.
    $a$, $e$, and `P.A.' denote the major axis, eccentricity and position angle of the elliptical disk model used for \msh, respectively.
    Results obtained with the 3D likelihood analysis approach are annotated with `3D'.
  }
\end{sidewaystable*}

\section{Additional maps and spectra for the 3D likelihood analysis}
\label{sec:appendix_template}

\begin{figure}
  \centering
  \includegraphics{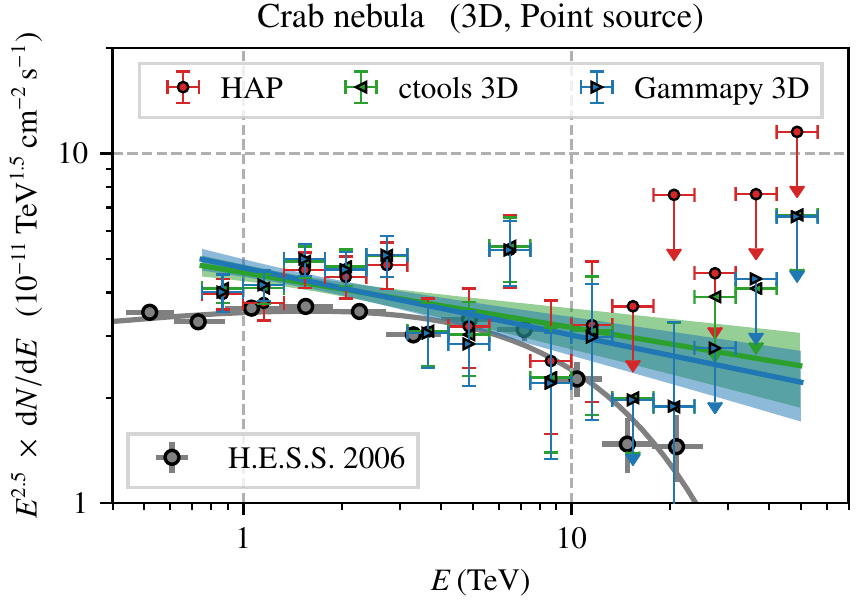}
  \caption{Comparison of spectra for the \crab.
  The spectra shown in green and blue were derived using a 3D likelihood analysis with \ct and \gp, respectively.
  The butterflies show the fitted power laws.
  The results are compared to those obtained with the reflected background method using the \hap software (in red).
  The published spectrum is taken from \citet{hesscrab2006}; it uses a power law with exponential cut-off as spectral model.}
  \label{fig:spectrum_3d_crab}
\end{figure}

\begin{figure}
  \centering
  \includegraphics{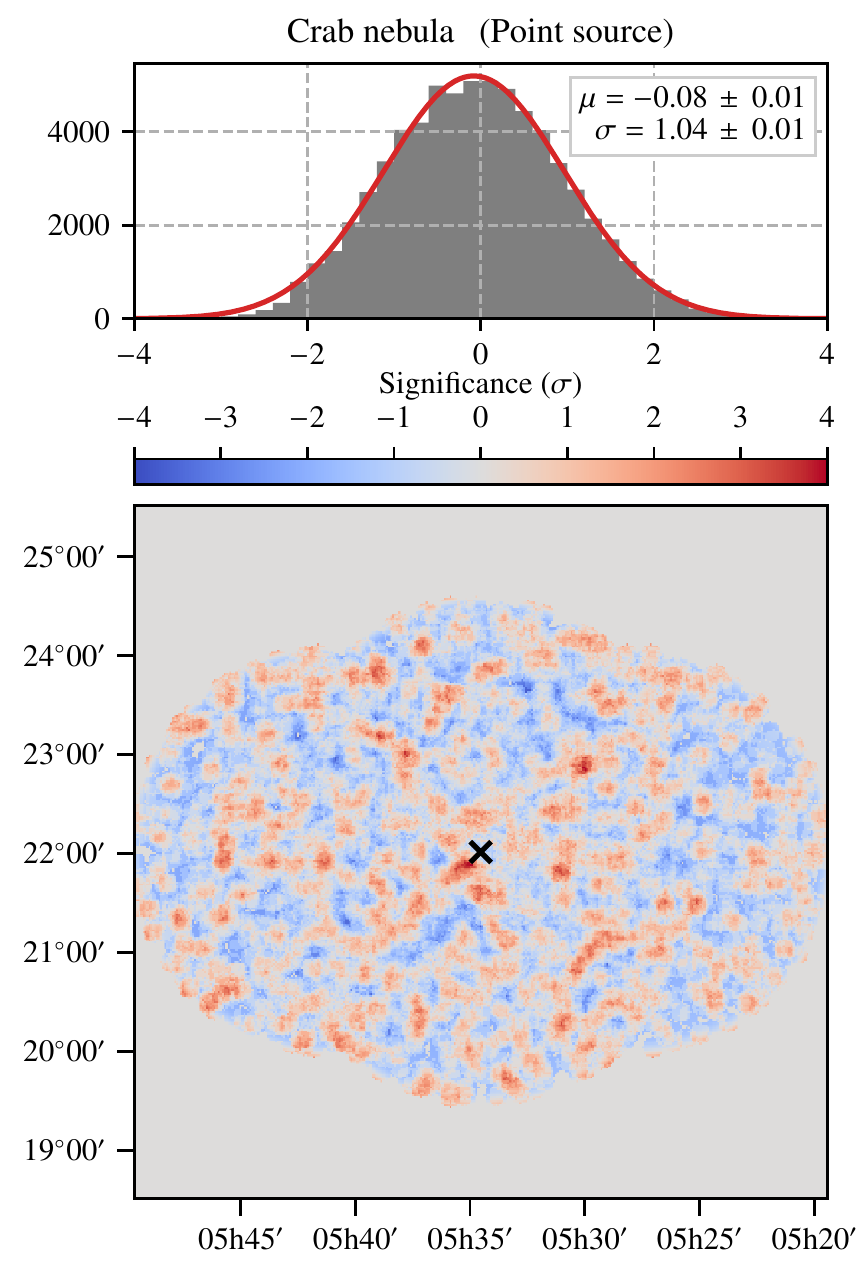}
  \caption{Residual significance map for the \crab, in equatorial coordinates (J2000).
  Results derived with the 3D likelihood analysis with \gp are displayed.
  We apply a convolution with a top-hat kernel of $0.1^\circ$ radius to reduce statistical fluctuations.
  The position of the \crab is indicated by the `$\times$'.
  The upper panel shows the distribution of significance values, together with the fit of a normal distribution (red line).}
  \label{fig:sign_map_3d_crab}
\end{figure}

We show the spectrum and significance map obtained with the 3D likelihood analysis for the \crab in Figs.~\ref{fig:spectrum_3d_crab} and~\ref{fig:sign_map_3d_crab}, respectively.

\begin{figure}
  \centering
  \includegraphics{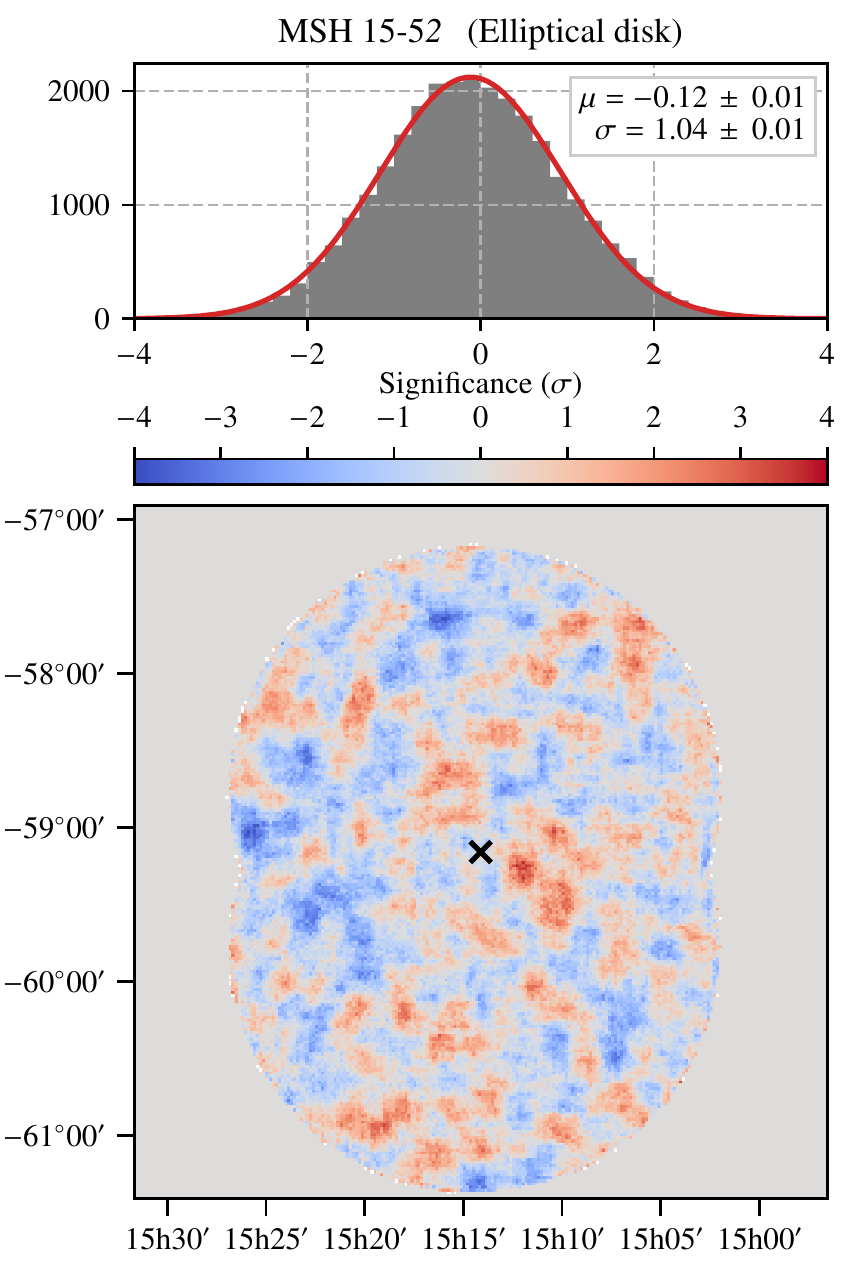}
  \caption{Residual significance map for \msh, in equatorial coordinates (J2000).
  Results derived with the 3D likelihood analysis with \ct using an elliptical disk as source model are displayed.
  We apply a convolution with a top-hat kernel of $0.1^\circ$ radius to reduce statistical fluctuations.
  The position of \msh is indicated by the `$\times$'.
  The upper panel shows the distribution of significance values, together with the fit of a normal distribution (red line).}
  \label{fig:sign_map_3d_msh1552}
\end{figure}

Figure~\ref{fig:sign_map_3d_msh1552} shows the significance map for the 3D likelihood analysis of \msh with an elliptical disk model.

\begin{figure*}
  \centering
  \includegraphics{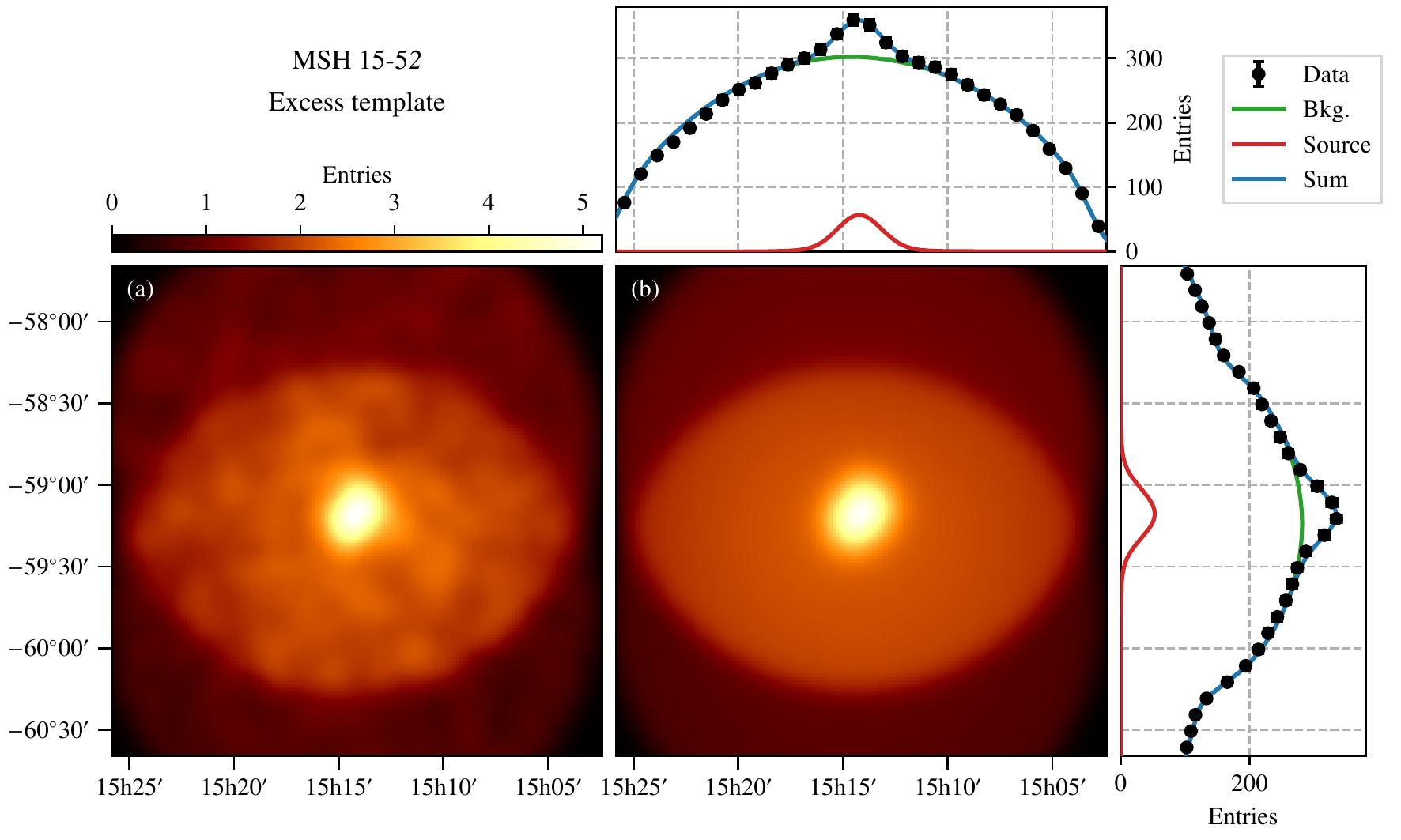}
  \caption{Counts map (a) and best-fit model map (b) for \msh, in equatorial coordinates (J2000).
  Results derived with the 3D likelihood analysis with \ct using an excess template as source model are displayed.
  The maps have been integrated over all energy bins contributing to the fit and smoothed using a Gaussian kernel with a width of $0.08^\circ$.
  The small panels show projections onto the two spatial axes; the separate components are indicated as well here.}
  \label{fig:data_model_map_3d_msh1552_excess}
\end{figure*}

\begin{figure}
  \centering
  \includegraphics{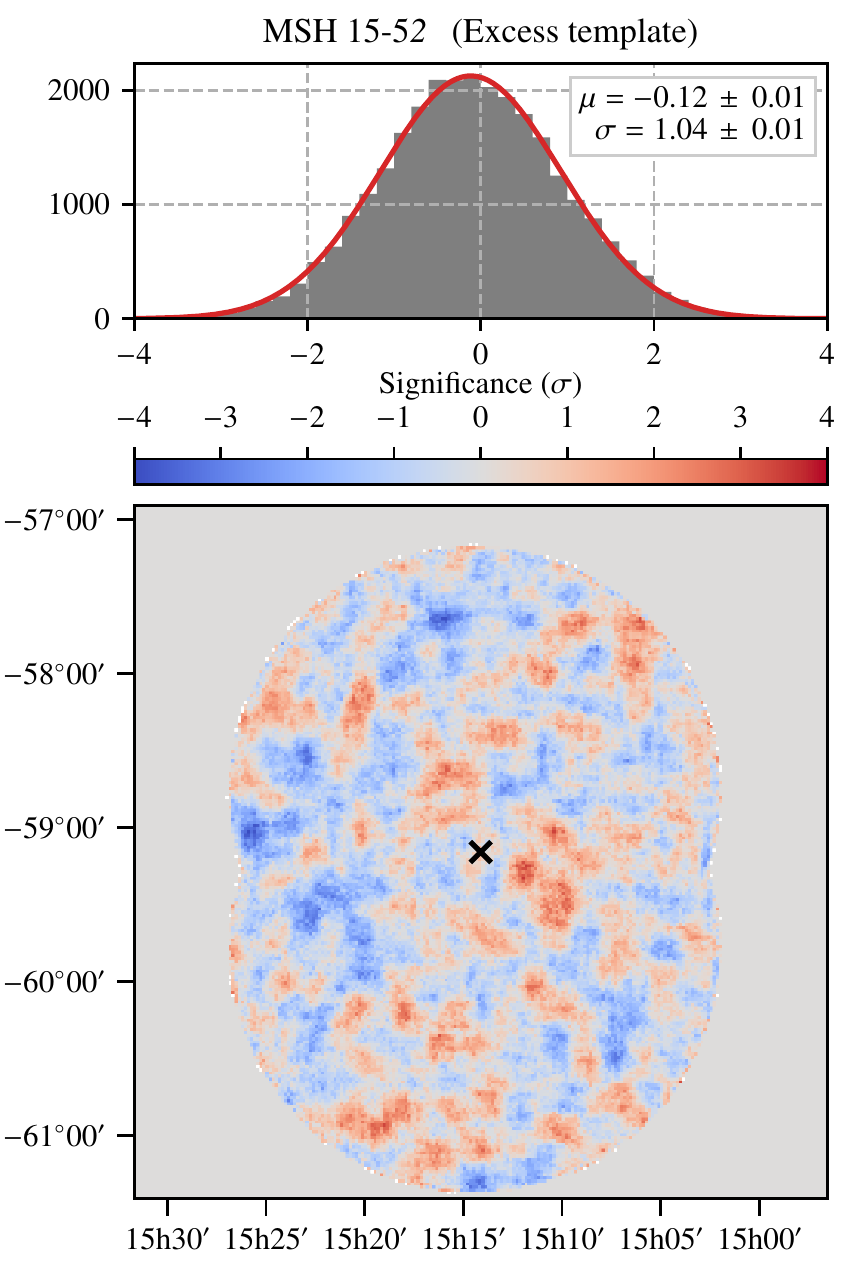}
  \caption{Residual significance map for \msh, in equatorial coordinates (J2000).
  Results derived with the 3D likelihood analysis with \ct using an excess template as source model are displayed.
  We apply a convolution with a top-hat kernel of $0.1^\circ$ radius to reduce statistical fluctuations.
  The position of \msh is indicated by the `$\times$'.
  The upper panel shows the distribution of significance values, together with the fit of a normal distribution (red line).}
  \label{fig:sign_map_3d_msh1552_excess}
\end{figure}

\begin{figure}
  \centering
  \includegraphics{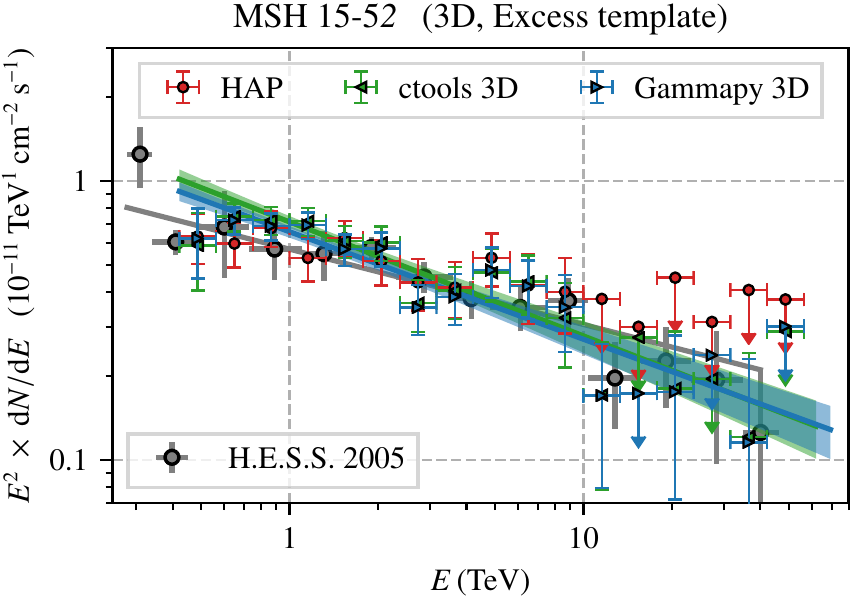}
  \caption{Comparison of spectra for \msh.
  The spectra shown in green and blue were derived using a 3D likelihood analysis with \ct and \gp, respectively, employing an excess template as source model.
  The butterflies show the fitted power laws.
  The results are compared to those obtained with the reflected background method using the \hap software (in red).
  The published spectrum is taken from \citet{hessmsh1552_2005}.}
  \label{fig:spectrum_3d_msh1552_excess}
\end{figure}

Figures~\ref{fig:data_model_map_3d_msh1552_excess}-\ref{fig:spectrum_3d_msh1552_excess} display the results of the 3D likelihood analysis of \msh with an excess template model.

\begin{figure*}
  \centering
  \includegraphics{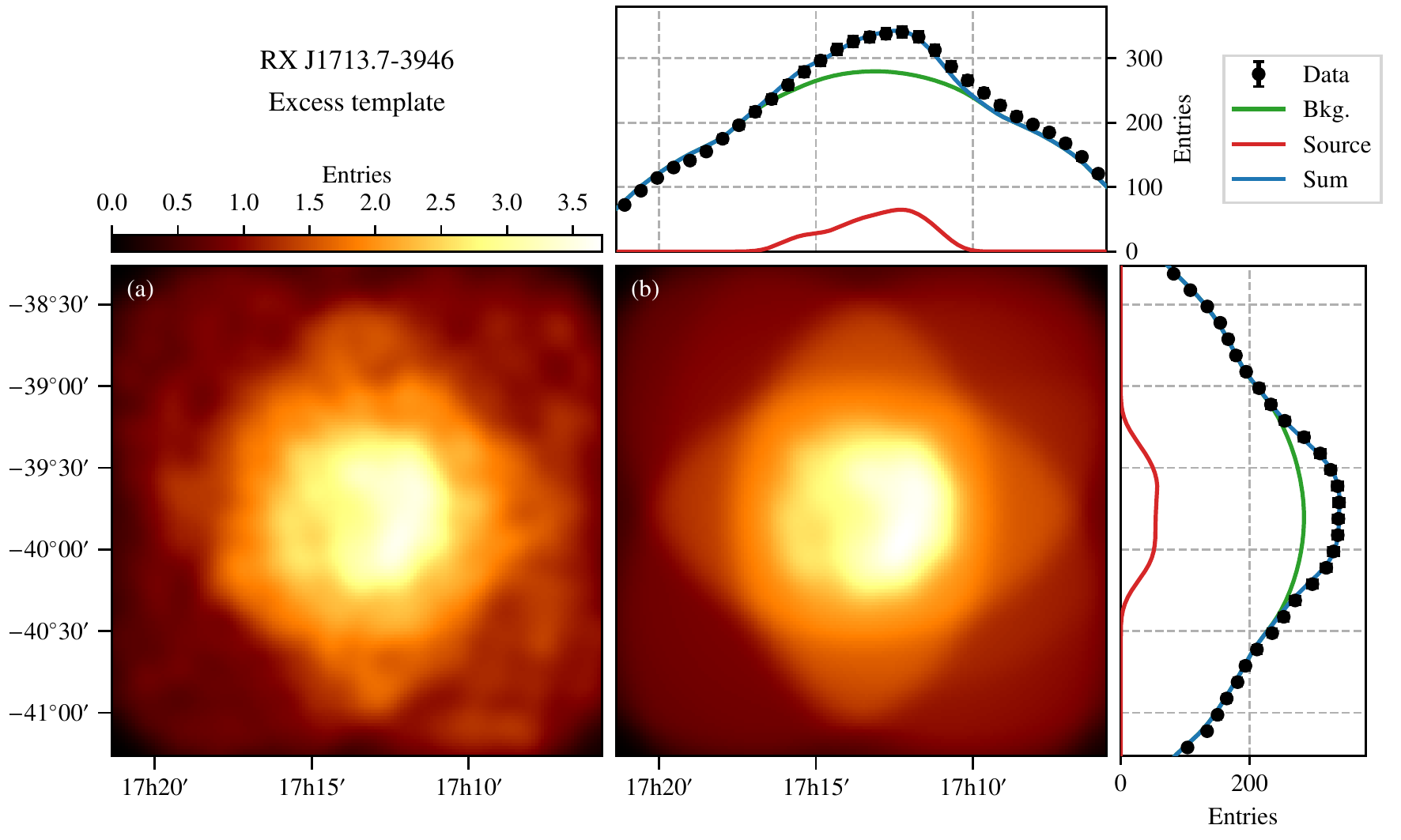}
  \caption{Counts map (a) and best-fit model map (b) for \rxj, in equatorial coordinates (J2000).
  Results derived with the 3D likelihood analysis with \gp using an excess template as source model are displayed.
  The maps have been integrated over all energy bins contributing to the fit and smoothed using a Gaussian kernel with a width of $0.08^\circ$.
  The small panels show projections onto the two spatial axes; the separate components are indicated as well here.}
  \label{fig:data_model_map_3d_rxj1713}
\end{figure*}

\begin{figure}
  \centering
  \includegraphics{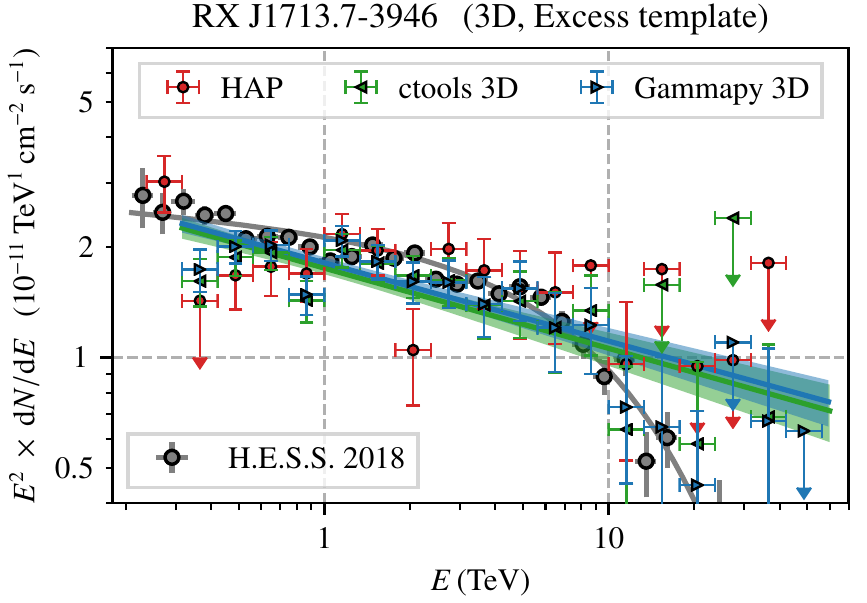}
  \caption{Comparison of spectra for \rxj.
  The spectra shown in green and blue were derived using a 3D likelihood analysis with \ct and \gp, respectively, employing an excess template as source model.
  The butterflies show the fitted power laws.
  The results are compared to those obtained with the reflected background method using the \hap software (in red).
  The published spectrum is taken from \citet{hessrxj1713_2018}; it uses a power law with exponential cut-off as spectral model.}
  \label{fig:spectrum_3d_rxj1713}
\end{figure}

Finally, we show the data and model maps and the spectrum derived with the 3D likelihood analysis for \rxj in Figs.~\ref{fig:data_model_map_3d_rxj1713} and~\ref{fig:spectrum_3d_rxj1713}, respectively.

\section[Supplementary material: background model templates]{Supplementary material:\\background model templates}
\label{sec:appendix_bkg_model}
We make available the three-dimensional templates for the residual cosmic-ray background that we derive in this paper for all observations that are part of the first public \hess test data release \citep[see][]{hesstestdata}.
We note that the usage of the public \hess test data set is subject to the terms of use that are distributed together with the data, in particular ``no scientific publications may be derived from the data''.

The material can be found at the following URL:\\
\url{https://github.com/lmohrmann/hess\_ost\_paper\_material}

\section{Supplementary material: machine-readable tables of spectral results}
\label{sec:appendix_result_tables}
We release as ASCII text files the results of all spectral fits carried out in this paper.
Both the results of the fitted power-law models as well as extracted spectral flux points are available for each of the analysis tools that we have used (i.e.\ the \hess-internal analysis software program \hap and the open-source packages \ct and \gp).

The material can be found at the following URL:\\
\url{https://github.com/lmohrmann/hess\_ost\_paper\_material}

\end{appendix}

\end{document}